\documentclass[fleqn,usenatbib]{mnras}

\usepackage{newtxtext,newtxmath}
\usepackage{graphicx}  

\usepackage[T1]{fontenc}
\usepackage{algorithm,algpseudocode}

\usepackage{siunitx}
\DeclareRobustCommand{\VAN}[3]{#2}
\let\VANthebibliography\thebibliography
\def\thebibliography{\DeclareRobustCommand{\VAN}[3]{##3}\VANthebibliography}

\usepackage{xcolor}
\usepackage{amsmath, amsfonts, bm} 	
\usepackage{hyperref}

\title{A $Z_1^2$ framework for rotational-parameter estimation and uncertainty quantification in high-energy pulsars}

\author[A.~Singhal et al.]{
Akshat Singhal,$^{1}$
Rohit Nair,$^{2,3}$
Devendra Sahu$^{4,5}$
Gayathri Raman$^{6}$
Suman Bala$^{6, 7}$
\\
$^{1}$Homi Bhabha Center for Science Education, Tata Institute of Fundamental Research, Mumbai - 400088, India\\
$^{2}$St. Xavier's College, Mumbai - 400001, India\\
$^{3}$Department of Physics, University of Mumbai\\
$^{4}$Center for Basic Sciences at Pt. Ravishankar Shukla University, Raipur - 492010, India\\
$^{5}$Indian Institute of Science Education and Research Bhopal, Bhopal - 462066, India \\
$^{6}$Indian Institute of Technology Bombay, Mumbai - 400076, India\\
$^{7}$USRA STI, Columbia, USA
}

\date{Accepted XXX. Received YYY; in original form ZZZ}

\pubyear{\the\year{}}

\begin{document}
\label{firstpage}
\pagerange{\pageref{firstpage}--\pageref{lastpage}}
\maketitle

\begin{abstract}
We present a $Z_1^2$-based framework for estimating the spin frequency and frequency derivative of high-energy pulsars from Poisson-limited photon event lists. The key point is that the width of a coherent detection peak is not, by itself, the statistical uncertainty on the recovered rotational parameters. We develop and compare three computationally efficient estimators: segmented frequency regression, a coherent derivative scan, and a localized two-dimensional coherent fit. For sinusoidal signals, we derive the local form of the $Z_1^2(f,\dot f)$ response and show that expressing the frequency at the midpoint of the observation removes the leading-order covariance between $f$ and $\dot f$. This gives simple uncertainty estimates in terms of the fitted peak amplitude and local widths, without requiring an exhaustive Monte Carlo simulation for each observation. We test these estimates with Monte Carlo simulations over a range of observing spans, signal strengths, grid resolutions, and good-time-interval structures, and show that the predicted uncertainties reproduce the run-to-run scatter of the recovered parameters in the tested regimes. We then apply the framework to \textit{AstroSat}/LAXPC event lists for the Crab pulsar, Swift~J0243.6+6124, and SAX~J1808.4--3658. The results provide a practical and statistically motivated route to rotational-parameter estimation for targeted high-energy pulsar searches.
\end{abstract}

\begin{keywords}
pulsars: general -- stars: neutron -- X-rays: stars -- methods: data analysis -- methods: statistical
\end{keywords}



\section{Introduction}

\subsection{Pulsar timing in the high-energy regime}

Neutron stars are compact remnants of massive stars and provide some of the most useful astrophysical laboratories for studying matter, gravity, and plasma processes under extreme conditions. Since their discovery as pulsating radio sources \citep{Hewish1968}, pulsars have remained central to a wide range of astrophysical studies. Their rotational stability allows precise timing measurements, which have been used to probe dense-matter physics \citep{Lattimer2004}, test general relativity in strong gravitational fields \citep{Kramer2006}, and, more recently, contribute to the detection of nanohertz gravitational waves through pulsar timing arrays \citep{Agazie2023,Antoniadis2023}.

A central task in pulsar timing is the measurement of the rotational phase model. To lower order, this model is described by the spin frequency $f$ and its first derivative $\dot f$. These parameters encode the secular evolution of the star and are used to infer standard derived quantities such as the characteristic age, surface magnetic-field strength, and spin-down luminosity \citep{Manchester1977,Lorimer2005}. For isolated rotation-powered pulsars, $\dot{f}$ is expected to be negative, reflecting the steady loss of rotational energy through electromagnetic and particle-wind braking. In accreting systems, the accretion flow can either spin up or spin down the neutron star, so $\dot{f}$ may be positive or negative depending on the instantaneous torque. 

In radio timing, pulsars are commonly observed with sufficient signal-to-noise ratio to form stable pulse profiles. Pulse times of arrival (TOAs) can then be measured by comparing the observed profile with a high signal-to-noise template \citep{Taylor1992,Lorimer2005}. The timing model is fitted to these TOAs, and the associated parameter uncertainties are usually obtained from a least-squares or generalized least-squares framework. In this setting, the measurement problem can often be reduced to fitting a sequence of TOAs with approximately Gaussian uncertainties.

The situation is different for many high-energy pulsar observations. Instruments such as the \textit{Fermi} Large Area Telescope \citep{Atwood2009}, the Neutron star Interior Composition Explorer \citep{Gendreau2016}, and \textit{AstroSat} \citep{2006AdSpR..38.2989A,2014SPIE.9144E..1SS} record individual photon arrival times. For faint sources, short observations, or restricted energy selections, the data can be sparse enough that reliable TOAs cannot be formed over short intervals. The analysis must then work directly with event arrival times, and the relevant statistical fluctuations are Poissonian rather than Gaussian \citep{Ray2011}.

\subsection{Period estimation from sparse photon arrival times}

For a photon detected at barycentre-corrected time $t_j$, the rotational phase for a trial timing solution is computed from a Taylor expansion about a reference epoch $t_0$,
\begin{align}
    \phi_j(f,\dot f)
    =
    f(t_j-t_0)
    +
    \frac{1}{2}\dot f(t_j-t_0)^2
    + \cdots
    \pmod{1}.
\end{align}
For the correct timing parameters, the phases trace the source pulse profile. For an incorrect solution, phase coherence is progressively lost and the phase distribution approaches that expected from noise.

Periodicity statistics measure this departure from phase uniformity. Epoch folding compares a binned phase profile with the expectation from a uniform distribution \citep{1978ApJ...224..953S,davies1990improved,Leahy1983b}, but its sensitivity depends on the binning choice. The $Z_n^2$ statistic instead uses the unbinned phases and measures the Fourier power contained in the first $n$ harmonics \citep{Rayleigh1919,Buccheri1983},
\begin{align}
    Z_n^2
    =
    \frac{2}{N}
    \sum_{k=1}^{n}
    \left[
    \left(\sum_{j=1}^{N}\cos 2\pi k\phi_j\right)^2
    +
    \left(\sum_{j=1}^{N}\sin 2\pi k\phi_j\right)^2
    \right],
\end{align}
where $N$ is the number of photons. The case $n=1$ corresponds to the Rayleigh statistic and is sensitive to the fundamental sinusoidal component of the pulse profile. Larger values of $n$ include harmonic power from sharper or multi-peaked profiles, but also increase the number of noise degrees of freedom. In the absence of a periodic signal, $Z_n^2$ follows a $\chi^2$ distribution with $2n$ degrees of freedom. The $H$-test addresses the choice of $n$ by selecting the number of harmonics adaptively, with a penalty for unnecessary terms \citep{busching2010h}.

In the present work, we limit the analytical development to the fundamental harmonic, $Z_1^2$. This choice is motivated by analytical simplicity. The $n=1$ case retains the essential structure of the coherent peak while avoiding the additional harmonic structure introduced for $n>1$. The behaviour of higher-order $Z_n^2$ statistics can be constructed from the corresponding harmonic contributions, as discussed in \citet{2023arXiv231106620S}. Thus, $Z_1^2$ provides a natural starting point for developing the parameter-uncertainty framework considered here.
\subsection{Spin-parameter recovery and uncertainty estimation}

In a targeted high-energy timing analysis, the periodicity statistic is evaluated over a restricted region of trial parameter space. If the spin evolution is approximated as constant over the observing window, the search produces a one-dimensional peak, $Z_1^2(f)$. If the first frequency derivative is included, the search produces a two-dimensional surface, $Z_1^2(f,\dot f)$. The coordinates of the `peak' define the recovered rotational parameters; hereafter, we denote recovered quantities with a subscript $r$, such as $f_r$ and $\dot f_r$.

The recovery of a maximum and the assignment of an uncertainty to that maximum are distinct problems. The width of the $Z_1^2$ peak is primarily set by the time span of the observation, with natural scales of order $1/T$ in frequency and $1/T^2$ in frequency derivative. The statistical scatter of the recovered maximum, however, also depends on the strength of the coherent signal. This dependence is already suggested by analytic treatments of sinusoidal signals, where the frequency uncertainty scales approximately as $\delta f \propto T^{-1}(Z_{1,\max}^2)^{-1/2}$ \citep[e.g.][]{Bretthorst1988,ransom2002fourier,chang}. Thus, the nominal Fourier resolution or width of the peak is not, by itself, a confidence interval on the recovered parameter.

This distinction becomes especially important when $f$ and $\dot f$ are fitted jointly. The two parameters can be correlated, and the uncertainty assigned to the frequency depends on the reference epoch at which the frequency is quoted. A frequency measured at the beginning of an observation, for example, need not have the same uncertainty as the frequency quoted near the middle of the observing window. Any uncertainty prescription for a two-dimensional search must therefore account for the local geometry of the $Z_1^2(f,\dot f)$ surface, not only for the width of one-dimensional slices.

Several approaches are commonly used to estimate timing uncertainties. When reliable pulse times of arrival are available, the timing parameters and their uncertainties can be obtained from a least-squares or generalized least-squares fit to the TOAs \citep[e.g.][]{Hobbs2006,Edwards2006,Verbiest2009}. In sparse high-energy data, however, individual TOAs may not be measurable. Likelihood-based analyses can instead fit the phase model directly to the photon arrival times \citep[e.g.][]{Ray2011,Kerr2011,Bretthorst1988}, but the resulting scalings with observation duration, peak strength, and search geometry are not always transparent.

A more empirical practice is to infer the uncertainty from the shape of the detection peak itself, for example by fitting an analytic profile, using a fixed drop in the statistic, or quoting a width such as the full width at half maximum \citep[e.g.][]{Bloomfield1976,Kovacs1980,Larsson1996,Leahy1987,2005ApJ...618..866N,2017NewA...56...94B}. Such measures describe the morphology of a given search surface, but they do not automatically give the sampling variance of the recovered maximum under repeated Poisson realizations.

Direct Monte Carlo simulations provide the cleanest frequentist benchmark, because they repeat the full search over many independent event lists with the same underlying source parameters. In practice, such simulations are often used as observation-specific checks or calibrations in timing and pulsation analyses, especially when the uncertainty is affected by the search procedure, pulse morphology, or source variability \citep[e.g.][]{Boldin2013,Raman2016,Raman2021}. Their computational cost, however, makes them impractical as the default uncertainty estimator for every observation.


The problem addressed in this work is therefore to connect the local structure of the $Z_1^2$ peak with the frequentist scatter of the recovered parameters in Poisson-limited data. This connection allows the uncertainty to be tied to the measured peak strength and search geometry, rather than to the grid spacing or to the peak width alone.

\subsection{Scope of this work}

This paper extends the $Z_1^2$-based uncertainty framework of \citet{2023arXiv231106620S} from the single-parameter frequency search to the two-parameter problem involving both $f$ and $\dot f$. Given a narrow-band search surface $Z_1^2(f,\dot f)$, we ask how the location of its maximum should be interpreted as a recovered timing solution, and how statistically meaningful confidence intervals can be assigned to that solution in the Poisson-limited regime.

The framework is intended for targeted searches rather than blind all-sky surveys. We assume that the source position is known, that the event times have been barycentre-corrected, and that the search is restricted to a small neighbourhood of an expected rotational solution. We further assume that the rotational evolution is stable over the analysed interval, with no unresolved glitches or strong timing noise, and that the mean count rate is approximately stable within the good time intervals.

Under these assumptions, we develop and compare three computationally efficient approaches for estimating $f$, $\dot f$, and their uncertainties. Their statistical behaviour is calibrated using Monte Carlo simulations and then evaluated on \textit{AstroSat}/LAXPC observations of the Crab pulsar, Swift~J0243.6+6124, and SAX~J1808.4--3658.

The remainder of this paper is organized as follows.
Section~\ref{sec:observations} describes the observations, source sample, and data reduction.
Section~\ref{sec:simulations} introduces the simulation framework.
Section~\ref{sec:methods} presents the three parameter-estimation methods.
Section~\ref{sec:validation} validates the uncertainty estimates using Monte Carlo simulations.
Section~\ref{sec:results} applies the methods to the AstroSat/LAXPC event lists.
Section~\ref{sec:discussion} discusses the applicability and limitations of the framework and summarizes the conclusions.

\section{Observations and Data Reduction}\label{sec:observations}

We test the proposed uncertainty-estimation framework using archival observations from the Large Area X-ray Proportional Counter (LAXPC) instrument onboard \textit{AstroSat} \citep{2006AdSpR..38.2989A,2014SPIE.9144E..1SS,2017ApJS..231...10A}. LAXPC consists of three co-aligned proportional counter units, LAXPC10, LAXPC20, and LAXPC30, operating over the 3--80~keV energy range with event-mode timing capability. Its large effective area and high time resolution make it well suited for testing timing methods on bright and moderately bright X-ray pulsars.

We use three sources chosen to sample different timing regimes: the Crab pulsar, an isolated rotation-powered pulsar used here as a high-SNR benchmark; SAX~J1808.4--3658, an accreting millisecond X-ray pulsar with coherent pulsations near 401~Hz; and Swift~J0243.6+6124, a Galactic ultraluminous X-ray pulsar that underwent strong spin evolution during its 2017 outburst. These systems are not analysed here to obtain new astrophysical constraints on their long-term evolution. Instead, they provide real event lists with different count rates, spin frequencies, frequency derivatives, and GTI structures against which the proposed uncertainty framework can be tested.

\begin{table}
\centering
\caption{
Summary of the \textit{AstroSat}/LAXPC observations used to test the timing-uncertainty framework.
The elapsed span is measured between the first and last photon in the screened event list.
The gap fraction denotes the percentage of this span excluded by good-time-interval (GTI) filtering, primarily due to Earth occultation, South Atlantic Anomaly passages, and other standard screening criteria.
}
\label{tab:obs_summary}
\setlength{\tabcolsep}{6pt}
\begin{tabular}{l c r r}
\hline
Source & ObsID & Observed time& Gap \\
&& (ks)& (\%)\\
\hline
Crab pulsar          & 9000001594 & 250 & 60 \\
Swift~J0243.6+6124  & 9000001590 &  50 & 25 \\
SAX~J1808.4--3658   & 9000003090 & 118 & 25 \\
\hline
\end{tabular}
\end{table}

\subsection{Event processing and filtering}

We downloaded Level-1 data from the Indian Space Science Data Centre archive\footnote{\url{https://astrobrowse.issdc.gov.in/astro_archive/archive/Home.jsp}} and processed the Event Analysis mode data using \textsc{LaxpcSoft} version 3.1.1, distributed by the \textit{AstroSat} Science Support Cell\footnote{\url{http://astrosat-ssc.iucaa.in}}. Standard event files were generated with \texttt{laxpc\_make\_event}. Good time intervals were produced using \texttt{laxpc\_make\_stdgti}, excluding intervals affected by Earth occultation and passages through the South Atlantic Anomaly. The standard detector dead-time correction was applied to the extracted products.

For each source, we applied energy and detector-layer selections chosen to improve the pulsation signal-to-noise ratio. The observation identifiers, elapsed spans, and GTI gap fractions are summarized in Table~\ref{tab:obs_summary}. Photon arrival times were barycentre-corrected using the source coordinates adopted for each target before evaluating the timing statistics.

\subsection{Source sample}\label{sec:sources}

\begin{table}
\centering
\caption{
Reference timing properties of the pulsars used as real-data test cases.
The values are representative literature measurements and are included only to indicate the approximate spin-frequency and frequency-derivative scales relevant for the narrow-band searches.
They are not intended as a homogeneous timing solution, since the quoted measurements correspond to different epochs, instruments, and analysis assumptions.
For SAX~J1808.4--3658, the quoted $\dot f$ range spans long-term spin-down estimates and outburst-local measurements.
References: (1) \citet{Lyne1993_CrabTiming}; 
(2) \citet{wijnands1998millisecond,chakrabarty1998two,2020ApJ...898...38B,sharma2023astrosat}; 
(3) \citet{2018AandA...613A..19D}.
}
\label{tab:source_timing_context}
\setlength{\tabcolsep}{5pt}
\begin{tabular}{l c c c}
\hline
Source & $f$ (Hz) & $\dot f$ (Hz s$^{-1}$) & Ref. \\
\hline
Crab pulsar          & $\sim 29.7$    & $\sim -3.7\times10^{-10}$   & 1 \\
SAX~J1808.4--3658   & $\sim 400.975$ & $\sim 10^{-15}$--$10^{-11}$ & 2 \\
Swift~J0243.6+6124  & $\sim 0.1015$  & $\sim 2.2\times10^{-10}$    & 3 \\
\hline
\end{tabular}
\end{table}

The three targets were selected to test the uncertainty-estimation framework across distinct timing regimes rather than to obtain new astrophysical constraints on their long-term evolution. They span a bright isolated rotation-powered pulsar, a rapidly rotating accretion-powered millisecond pulsar, and a slowly rotating ultraluminous X-ray pulsar with strong spin evolution. Approximate reference timing parameters from the literature are listed in Table~\ref{tab:source_timing_context}; these values are used only to define the relevant timing scale and to motivate the narrow-band search ranges.

The Crab pulsar provides a bright and rotationally stable benchmark. Its high X-ray count rate allows the behaviour of the $Z_1^2$ peak and the associated uncertainty estimates to be tested in a regime where the periodic signal is strongly detected. The pulsar has a period of approximately 33~ms and a well-measured secular spin-down, making it a useful reference source for validating whether the recovered frequency and frequency derivative are stable over short analysis segments.

SAX~J1808.4--3658 is the prototype accreting millisecond X-ray pulsar, with coherent pulsations near 401~Hz in a compact binary system \citep{wijnands1998millisecond,chakrabarty1998two}. It provides a test case at much higher spin frequency than the Crab. In such accretion-powered systems, the measured frequency derivative can be affected by accretion torques, orbital corrections, and timing noise, making robust uncertainty estimation especially important.

Swift~J0243.6+6124 is the first known Galactic ultraluminous X-ray pulsar \citep{kennea2017swift,cenko2017grb}. During its 2017 outburst it showed coherent pulsations near 9.86~s and rapid spin evolution \citep{jenke2017fermi,jaisawal2018understanding,beri2021astrosat}. It therefore provides a complementary test case in a slow accretion-powered pulsar undergoing strong spin evolution, where measuring $\dot f$ and its uncertainty is physically important.

\section{Simulation Framework}\label{sec:simulations}

We use controlled simulations to calibrate the behaviour of the rotational-parameter estimators. Since the injected parameters are known, the recovered values from repeated Poisson realisations can be compared directly with the true values and with the proposed uncertainty estimates.

Photon arrivals are modelled as an inhomogeneous Poisson process with time-dependent rate
\begin{align}
    y(t)
    =
    a
    +
    b\sin\left[
    2\pi\left(
    f_0 t + \frac{1}{2}\dot f_0 t^2
    \right)
    + \phi
    \right],
\end{align}
where $a$ is the mean count rate, $b$ is the modulation amplitude, $\phi$ is the initial phase, and $(f_0,\dot f_0)$ are the injected spin frequency and frequency derivative at the reference epoch $t=0$. We require $a>|b|$ so that the rate remains positive. Higher-order frequency derivatives are neglected in the simulations presented here.

\begin{figure}
    \centering
    \includegraphics[width=0.80\linewidth]{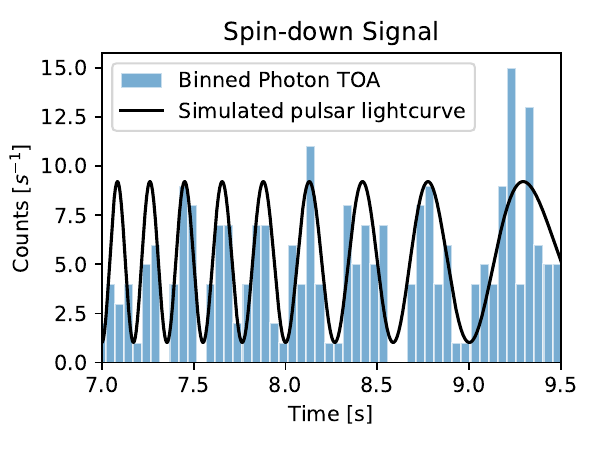}
    \caption{
    Schematic simulated photon arrival series for a sinusoidal source with an intentionally large spin-down term. 
    The underlying rate is $y(t)=100+80\sin[2\pi(20t-t^2)]$, corresponding to $f_0=20$~Hz and $\dot f_0=-2$~Hz~s$^{-1}$ at the reference epoch $t=0$. 
    The event times are binned at 0.05~s over an observation time of $T=9.5$~s. 
    These parameters are unphysical for pulsars and are used only to make the effect of the frequency derivative visually apparent.
    }
    \label{fig:simulated_spindown_signal}
\end{figure}

The frequency derivative term represents a secular change in the instantaneous spin frequency. It may correspond to spin-down, as in isolated rotation-powered pulsars, or spin-up, as in accretion-powered systems. Figure~\ref{fig:simulated_spindown_signal} shows a deliberately exaggerated example in which the changing pulse separation is visible by eye. The parameters in this figure are not intended to represent a physical pulsar; they are used only to illustrate how a non-zero $\dot f_0$ appears in an event stream.

\begin{figure}
    \centering
    \includegraphics[width=0.70\linewidth]{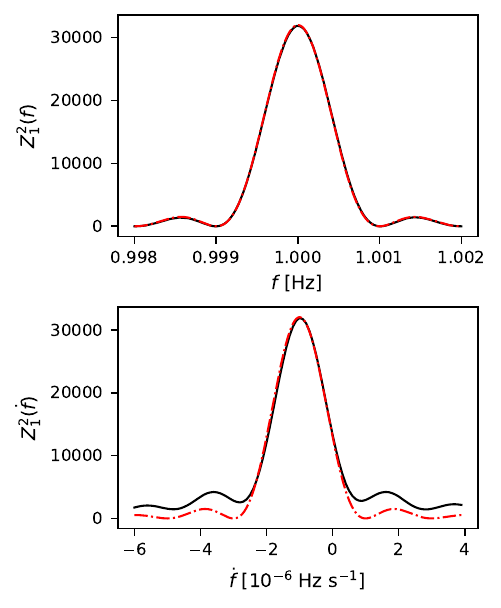}
    \caption{
    One-dimensional slices through the $Z_1^2$ surface for a high signal-to-noise simulated signal with $a=100$, $b=80$, $T=1000$~s, $f_0=1$~Hz, and $\dot f_0=-10^{-6}$~Hz~s$^{-1}$. 
    The upper panel shows $Z_1^2(f,\dot f=\dot f_0)$ and the lower panel shows $Z_1^2(f=f_0,\dot f)$. 
    The central part of each peak is well approximated by a local $\mathrm{sinc}^2$ profile, motivating the use of peak amplitude and peak width in the uncertainty estimates.
    }
    \label{fig:z1_sinc_slices}
\end{figure}

For each realisation, photon arrival times are drawn from the rate model and analysed using the same search procedure as for the observed event lists. The resulting recovered parameters define empirical distributions of $f_r$ and $\dot f_r$, from which we measure $\sigma(f_r)$ and $\sigma(\dot f_r)$.

\subsection{Search grid and local peak structure}

For each simulated or observed event list, we evaluate $Z_1^2$ over a narrow grid of trial rotational parameters. The search is centred on the injected values for simulations and on a reference rotational solution for real data. The natural scale of the coherent response is set by the observation time $T$: the frequency width is of order $1/T$, while the frequency-derivative width is of order $1/T^2$. The search grid is chosen to oversample these scales. In the simulations, the injected parameters are also given a random sub-bin offset so that they do not lie exactly on a grid point.

The statistic is computed using the implementation in \texttt{Stingray} \citep{huppenkothen2019stingray}. Unless stated otherwise, all simulations in this section use the fundamental harmonic, $n=1$, consistent with the analytical framework developed in this work.

Near its maximum, a one-dimensional slice of the $Z_1^2$ peak for a sinusoidal signal is well described by a local $\mathrm{sinc}^2$ profile,
\begin{align}
    Z(x)
    =
    A\,
    \mathrm{sinc}^2
    \left(
    \frac{x-x_r}{W_x}
    \right),
\end{align}
where $\mathrm{sinc}(u)\equiv \sin u/u$. Here $x$ denotes the searched parameter, $x_r$ is the recovered peak location, $A$ is the peak amplitude, and $W_x$ is the characteristic peak width in that direction. Hereafter, the subscript $r$ denotes a recovered quantity.

For an uninterrupted observation of duration $T$, the natural width scales are
\begin{align}
    W_f = \frac{1}{\pi T},
    \qquad
    W_{\mathrm{fd}} = \frac{2}{\pi T^2},
\end{align}
for the frequency and frequency-derivative directions, respectively. The first of these corresponds to the usual local width of the $Z_1^2(f)$ response, while the second gives the corresponding scale for a one-dimensional $\dot f$ slice near the central lobe. These widths are used both to interpret the peak morphology and to provide stable initial guesses for the local curve fits.

The quantities $A$, $W_f$, and $W_{\mathrm{fd}}$ provide the local information from which the uncertainty estimates are constructed. The widths describe the scale over which phase coherence is lost, while the peak amplitude measures the strength of the coherent signal relative to Poisson fluctuations.

Figure~\ref{fig:z1_sinc_slices} illustrates this local peak structure for a high signal-to-noise simulated event list. The frequency slice follows the expected $\mathrm{sinc}^2$ morphology closely. The frequency-derivative slice is less symmetric over the full plotted range because it is a slice through a two-dimensional coherent surface, but its central region near the maximum remains well described by the same local peak model.

\begin{figure}
    \centering
    \includegraphics[width=0.70\linewidth]{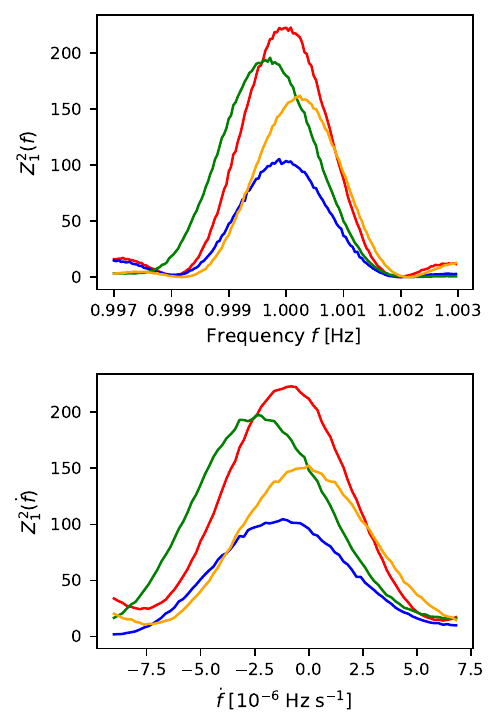}
    \caption{
    Variation of the $Z_1^2$ profiles across independent Poisson realisations of the same faint simulated source. 
    All realisations use identical injected parameters, $a=100$, $b=8$, $T=500$~s, $f_0=1$~Hz, and $\dot f_0=-10^{-6}$~Hz~s$^{-1}$. 
    Poisson fluctuations change both the peak amplitude and the recovered peak location. 
    The scatter in peak location illustrates why the macroscopic peak width alone is not a statistical uncertainty on the recovered rotational parameter.
    }
    \label{fig:z1_realization_scatter}
\end{figure}

A single event list gives only one realisation of the recovered peak. In Poisson-limited data, repeated observations of the same source would produce different photon arrival times and therefore slightly different $Z_1^2$ surfaces. Figure~\ref{fig:z1_realization_scatter} shows this effect for independent realisations of the same injected signal. Both the peak amplitude and the peak location vary from one realisation to another.

The methods developed in the next section use the measured peak structure to estimate $f_r$, $\dot f_r$, $\sigma(f_r)$, and $\sigma(\dot f_r)$. The Monte Carlo distributions from the simulations provide the benchmark against which those estimates are tested.

\section{Parameter-estimation Methods}\label{sec:methods}

The simulations in Section~\ref{sec:simulations} provide event lists for which the injected rotational parameters are known. We now describe the procedures used to recover $f$, $\dot f$, and their uncertainties from the measured $Z_1^2$ structure. The methods are designed for targeted, narrow-band searches in which the source position is known, the photon arrival times have been barycentre-corrected, and approximate rotational parameters are available from previous measurements or from a preliminary search.

The three approaches use different amounts of coherent information from the observation. The first approach divides the observation into shorter intervals, measures a local frequency in each interval, and obtains $\dot f$ from their time evolution. The second approach keeps the full observation coherent, but scans over trial values of $\dot f$ and monitors the recovered peak amplitude. The third approach performs a localized coherent search over both $f$ and $\dot f$ simultaneously. These approaches are complementary: the segmented approach is naturally robust to data gaps, the derivative scan efficiently refines the spin evolution, and the localized two-dimensional search gives the most direct joint estimate once the search region is sufficiently small.

\subsection{Method I: Segmented frequency regression}\label{sec:segmented_frequency_regression}

Segmented frequency regression estimates the frequency derivative by first measuring the local spin frequency in shorter intervals. The idea is closely related to time-resolved timing analyses in which local phase, frequency, or time-of-arrival measurements are fitted with a rotational model \citep[e.g.][]{Hobbs2006,Edwards2006,Verbiest2009}. The distinction here is that each local frequency is obtained directly from the $Z_1^2(f)$ peak of a Poisson-limited event list, and its uncertainty is assigned from the peak amplitude and width.

Consider an observation of duration $T$ divided into non-overlapping intervals of duration $t_{\mathrm{int}}$. Within the $i$-th interval, centred at time $t_i$, the signal is treated as approximately monochromatic. This is a local approximation: it is valid only if the phase drift caused by the frequency derivative within the interval remains sufficiently small. A useful order-of-magnitude condition is
\begin{align}
    \Delta \Phi_i
    \simeq
    \frac{1}{2}|\dot f|t_{\mathrm{int}}^2
    \lesssim
    1 .
\end{align}
At the same time, the interval must be long enough to contain sufficient photons for a well-localized $Z_1^2(f)$ peak. Thus, \(t_{\mathrm{int}}\) is not arbitrary: short intervals are limited by photon statistics, while long intervals are limited by intra-segment phase drift.

For each populated interval, we compute a one-dimensional $Z_1^2(f)$ curve over a narrow frequency range centred on the expected local frequency. The central peak is fitted with the same local $\mathrm{sinc}^2$ form introduced in Section~\ref{sec:simulations},
\begin{align}
    Z_i(f)
    =
    A_i\,
    \mathrm{sinc}^2
    \left(
    \frac{f-f_{r,i}}{W_{f,i}}
    \right),
\end{align}
where $f_{r,i}$ is the recovered frequency in the $i$-th interval, $A_i$ is the fitted peak amplitude, and $W_{f,i}$ is the fitted peak width. The uncertainty on the local frequency estimate is then assigned using the single-parameter relation
\begin{align}
    \sigma_{f,i}
    \simeq
    \frac{\sqrt{3}\,W_{f,i}}{\sqrt{A_i}},
\end{align}
following the $Z_1^2(f)$ uncertainty framework of \citet{2023arXiv231106620S}.

The set of local frequency measurements is then fitted as a function of interval midpoint time. We use the weighted linear model
\begin{align}
    f_{r,i}
    =
    f_{{\rm ref},r}
    +
    \dot f_r\,(t_i-t_{\rm ref})
    +
    \epsilon_i,
\end{align}
where the weights are $1/\sigma_{f,i}^2$. The slope gives the recovered frequency derivative $\dot f_r$, while the intercept gives the recovered frequency at the chosen reference epoch $t_{\rm ref}$. Choosing $t_{\rm ref}$ close to the middle of the observation reduces the covariance between the fitted intercept and slope.

\begin{algorithm}
\caption{Segmented frequency regression}\label{alg:segmented_frequency_regression}
\begin{algorithmic}[1]
\State Choose a segment duration $t_{\mathrm{int}}$ and divide the observation into non-overlapping intervals.
\State Discard intervals containing no usable events after GTI filtering.
\For{each populated interval $i$}
    \State Estimate the expected local frequency from the reference rotational parameters.
    \State Compute $Z_1^2(f)$ over a narrow frequency grid around this estimate.
    \State Isolate the central peak region around the primary maximum.
    \State Fit the local $\mathrm{sinc}^2$ profile to obtain $f_{r,i}$, $A_i$, and $W_{f,i}$.
    \State Assign the local uncertainty $\sigma_{f,i}=\sqrt{3}W_{f,i}/\sqrt{A_i}$.
\EndFor
\State Fit $f_{r,i}$ against $t_i$ using weighted linear regression.
\State Take the fitted slope as the recovered frequency derivative $\dot f_r$.
\end{algorithmic}
\end{algorithm}

\begin{figure*}
    \centering
    \includegraphics[width=0.8\textwidth]{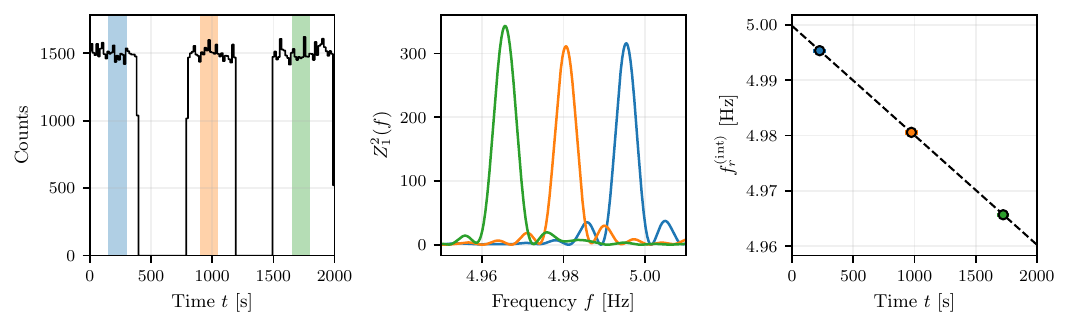}
    \caption{
    Illustration of segmented frequency regression for a simulated spin-down signal with data gaps.
    Left: binned event counts over the observation, with the three selected intervals highlighted.
    Middle: independent $Z_1^2(f)$ profiles computed in the selected intervals.
    Right: interval-wise recovered frequencies, $f_r^{(\mathrm{int})}$, plotted against interval midpoint time. 
    The weighted linear fit gives the recovered frequency derivative $\dot f_r$.
    The simulation uses $a=100$, $b=20$, $f_0=5.0~\mathrm{Hz}$, $\dot f_0=-2\times10^{-5}~\mathrm{Hz\,s^{-1}}$, $T=2000~\mathrm{s}$, and $t_{\rm int}=150~\mathrm{s}$.
    }
    \label{fig:spindown_analysis}
\end{figure*}

\begin{figure*}
    \centering
    \includegraphics[width=0.8\textwidth]{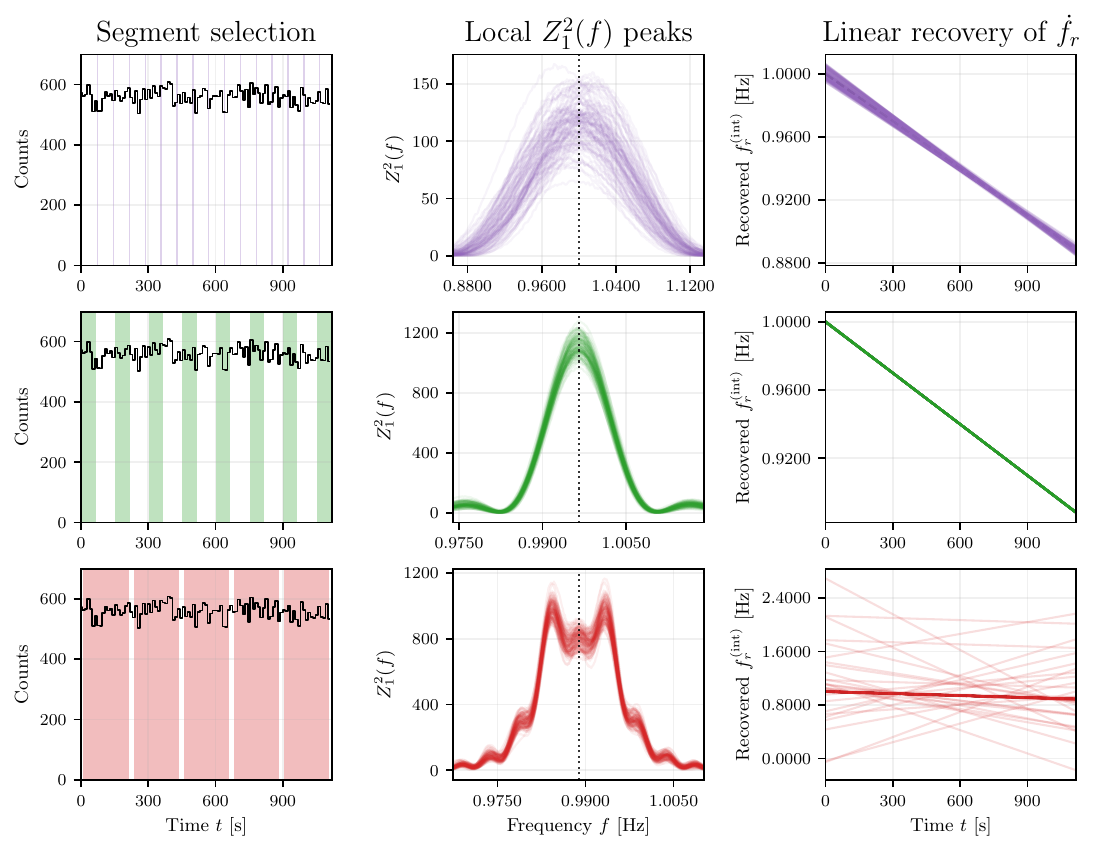}
    \caption{
    Effect of segment duration on the segmented frequency-regression method.
    Rows show three choices of $t_{\rm int}$: short segments (top), an intermediate choice (middle), and long segments (bottom).
    Left: segment selection, with the analysed time intervals shaded. For very short segments, only a subset of intervals is shaded for visual clarity.
    Middle: representative $Z_1^2(f)$ profiles from independent Poisson realisations for the first analysed interval. The dotted vertical line marks the expected local frequency.
    Right: recovered interval-wise frequencies, $f_r^{(\mathrm{int})}$, fitted as a function of interval midpoint time.
    Short segments are limited by photon statistics and produce broad, noisy peaks. Long segments suffer from intra-segment phase drift, which distorts the local peak and biases the frequency recovery. The intermediate regime balances these effects and gives the most stable recovery of $\dot f_r$.
    }
    \label{fig:method1_tint_regimes}
\end{figure*}

This approach has two practical advantages. First, it replaces a dense two-dimensional search over $(f,\dot f)$ with a set of one-dimensional frequency searches followed by a weighted regression. Secondly, it handles GTI gaps naturally: an interval without usable events simply contributes no local frequency measurement. The price paid is that the method uses only locally coherent information. It is therefore less statistically efficient than a fully coherent two-dimensional search when the observation is contiguous and the search region is already small.

\subsubsection{Choice of segment duration}\label{sec:segmented_tint_choice}

The segment duration controls the main bias--variance trade-off of this method. If $t_{\mathrm{int}}$ is too small, each interval contains too few photons, the local $Z_1^2(f)$ peak is broad or poorly defined, and the uncertainties $\sigma_{f,i}$ become large. If $t_{\mathrm{int}}$ is too large, the approximation of a constant frequency within the interval breaks down and the local peak can become distorted or split.

Figure~\ref{fig:method1_tint_regimes} illustrates these regimes using Monte Carlo realisations of the same injected signal. Very short intervals are dominated by photon statistics, very long intervals are affected by phase drift within each interval, and an intermediate range gives stable local frequency recovery. In the simulations considered here, useful segment durations are empirically found to lie approximately in the range
\begin{align}
    0.2
    \lesssim
    |\dot f|\,t_{\mathrm{int}}^2
    \lesssim
    3 , \label{eq:sweet}
\end{align}
provided that the local $Z_1^2(f)$ peaks are sufficiently strong and isolated. This condition should be interpreted as an operational guide rather than as a sharp mathematical boundary. The lower side of the range is set mainly by photon statistics, while the upper side is set by intra-segment phase drift and peak distortion.

The quantitative dependence of $\sigma(\dot f_r)$ on $t_{\mathrm{int}}$, the number of usable intervals $N_{\rm seg}$, and the presence of GTI gaps is examined in Section~\ref{sec:validation}. Those simulations are used to determine when the segmented approach is adequate on its own and when it is better used only as a preliminary step before a coherent derivative scan or a localized two-dimensional search.

\subsection{Method II: Coherent derivative scan}\label{sec:coherent_derivative_scan}

Segmented frequency regression is robust to gaps, but it sacrifices coherent signal power by analysing the observation in shorter intervals. When a sufficiently long portion of the event list can be treated coherently, the frequency derivative can instead be refined by scanning over trial values of $\dot f$ while retaining the full observation time.

For each trial value, denoted $\dot f_{\rm trial}$, we compute the photon phases using that frequency derivative and evaluate a one-dimensional $Z_1^2(f)$ curve over a local frequency grid. The central peak of this curve is fitted with the $\mathrm{sinc}^2$ form defined in Section~\ref{sec:simulations}. From this fit we record three quantities as functions of $\dot f_{\rm trial}$: the peak amplitude $A_{\rm pk}$, the peak frequency $f_{\rm pk}$, and the fitted frequency width $W_{\rm pk}$.

\begin{figure}
    \centering
    \includegraphics[width=0.8\columnwidth]{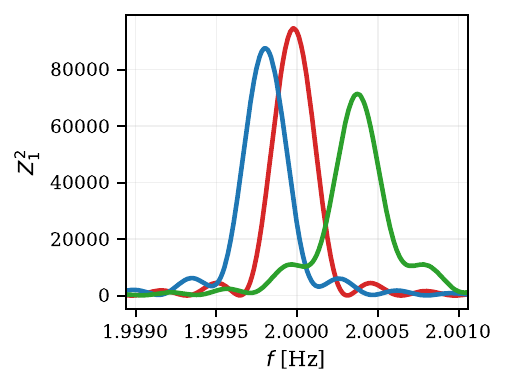}
    \caption{
    One-dimensional $Z_1^2(f)$ profiles obtained from the same simulated event list for three selected trial values of the frequency derivative, $\dot f_{\rm trial}$.
    The simulated source parameters are $a=100$, $b=80$, $f_0=2$~Hz, $\dot f_0=-3\times10^{-5}$~Hz~s$^{-1}$, and $T=3000$~s.
    The blue, red, and green curves correspond approximately to $\dot f_{\rm trial}=-3.025\times10^{-5}$, $-3.002\times10^{-5}$, and $-2.986\times10^{-5}$~Hz~s$^{-1}$, respectively.
    The profile closest to the coherent solution reaches the largest peak amplitude, whereas offset trial derivatives both shift the peak location in frequency and reduce the recovered power.
    }
    \label{fig:method2_profiles}
\end{figure}

\begin{figure}
    \centering
    \includegraphics[width=0.8\columnwidth]{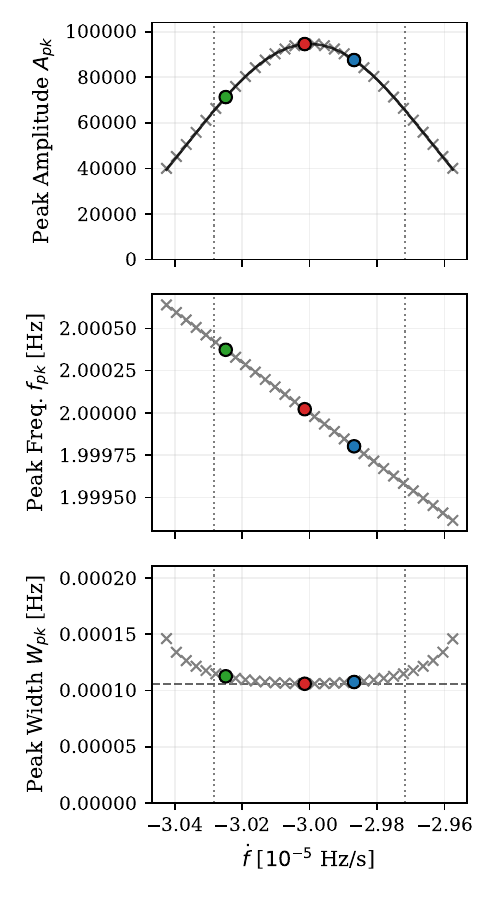}
    \caption{
    Fitted peak properties obtained from the coherent derivative scan.
    Each point is obtained by fitting the local $Z_1^2(f)$ peak at a fixed trial value of $\dot f_{\rm trial}$.
    The coloured markers correspond to the same three trial derivatives shown in Fig.~\ref{fig:method2_profiles}.
    Top: fitted peak amplitude $A_{\rm pk}$ as a function of $\dot f_{\rm trial}$. The central response is well described by a local $\mathrm{sinc}^2$-like envelope, whose maximum gives the recovered frequency derivative $\dot f_r$.
    Middle: fitted peak frequency $f_{\rm pk}$, which varies approximately linearly with $\dot f_{\rm trial}$ because a change in the trial frequency derivative can be partly compensated by a shift in frequency.
    Bottom: fitted peak width $W_{\rm pk}$, which remains approximately constant near the coherent solution.
    The vertical dotted lines mark the approximate central region $\dot f_r\pm\sqrt{15}\,W_{\mathrm{fd}}$, where $W_{\mathrm{fd}}=2/(\pi T^2)$.
    }
    \label{fig:method2_peak_properties}
\end{figure}

If $\dot f_{\rm trial}$ is close to the coherent solution, the accumulated phase drift is corrected and the recovered $Z_1^2(f)$ peak has a large amplitude. If $\dot f_{\rm trial}$ is offset from this value, phase coherence is partly lost over the observation and the peak amplitude decreases. Thus, the function $A_{\rm pk}(\dot f_{\rm trial})$ forms a coherent response envelope whose maximum gives the recovered frequency derivative. Near its central lobe, we model this envelope as
\begin{align}
    A_{\rm pk}(\dot f_{\rm trial})
    =
    A_0\,
    \mathrm{sinc}^2
    \left(
    \frac{\dot f_{\rm trial}-\dot f_r}{W_A}
    \right),
\end{align}
where $\dot f_r$ is the recovered frequency derivative and $W_A$ is the width of the amplitude response. This width is distinct from the direct $\dot f$-slice width $W_{\mathrm{fd}}$ introduced in Section~\ref{sec:simulations}. For an uninterrupted observation, the amplitude envelope is broader, with a useful initial scale
\begin{align}
    W_A
    \simeq
    \sqrt{15}\,W_{\mathrm{fd}}
    =
    \frac{2\sqrt{15}}{\pi T^2}.
\end{align}
In practice this scale is used as an initial estimate for the fit; the recovered centroid $\dot f_r$ is not sensitive to small changes in this starting width when the central lobe is well sampled.

The fitted peak frequency $f_{\rm pk}(\dot f_{\rm trial})$ provides a useful diagnostic of the covariance between frequency and frequency derivative. An error in the trial $\dot f$ can be partly compensated by shifting the recovered frequency. For an observation whose reference epoch is at the start of the interval, the approximately conserved combination is the midpoint frequency,
\begin{align}
    f_{\rm pk}(\dot f_{\rm trial})
    +
    \frac{T}{2}\dot f_{\rm trial}
    \simeq
    f_{\rm mid}.
\end{align}
Equivalently,
\begin{align}
    f_{\rm pk}(\dot f_{\rm trial})
    \simeq
    f_{\rm mid}
    -
    \frac{T}{2}\dot f_{\rm trial}.
\end{align}
This linear ridge anticipates the coordinate transformation used in the localized two-dimensional search described in Section~\ref{sec:localized_2d_search}.

Once $\dot f_r$ has been obtained from the amplitude envelope, we evaluate $Z_1^2(f)$ at $\dot f=\dot f_r$ and fit the resulting one-dimensional frequency peak to obtain the recovered frequency $f_r$. The corresponding midpoint frequency is
\begin{align}
    f_{{\rm mid},r}
    =
    f_r
    +
    \frac{1}{2}\dot f_r T,
\end{align}
for a reference epoch at the start of the observation. As shown later, this midpoint frequency is less correlated with $\dot f_r$ than the frequency quoted at the start epoch.

\begin{algorithm}
\caption{Coherent derivative scan}\label{alg:coherent_derivative_scan}
\begin{algorithmic}[1]
\State Define a grid of trial frequency derivatives, $\dot f_{\rm trial}$, centred approximately around preliminary estimate.
\For{each $\dot f_{\rm trial}$}
    \State Compute photon phases using $\dot f_{\rm trial}$.
    \State Evaluate $Z_1^2(f)$ over a local frequency grid.
    \State Fit the central $Z_1^2(f)$ peak with the local $\mathrm{sinc}^2$ model.
    \State Record $A_{\rm pk}(\dot f_{\rm trial})$, $f_{\rm pk}(\dot f_{\rm trial})$, and $W_{\rm pk}(\dot f_{\rm trial})$.
\EndFor
\State Fit $A_{\rm pk}(\dot f_{\rm trial})$ with a local $\mathrm{sinc}^2$ envelope.
\State Take the centroid of this envelope as $\dot f_r$.
\State Recompute $Z_1^2(f)$ at $\dot f=\dot f_r$ and fit the resulting peak to obtain $f_r$.
\end{algorithmic}
\end{algorithm}

The coherent derivative scan is most useful as an intermediate refinement step. It uses the full observation and is therefore more sensitive than segmented regression, but it avoids the cost of a dense two-dimensional search. Its main limitation is that the observation must be long enough for incorrect trial derivatives to produce a measurable loss of coherent power. The uncertainty scaling and the correlation between $f_r$ and $\dot f_r$ are examined with Monte Carlo simulations in Section~\ref{sec:validation}.

\subsection{Method III: Local two-dimensional coherent fitting}
\label{sec:localized_2d_search}

The coherent derivative scan uses the full observation time, but for every trial value of $\dot f_{\rm trial}$ it still requires a one-dimensional search over frequency. Once the approximate location of the coherent peak is known, a more direct strategy is to evaluate $Z_1^2(f,\dot f)$ on a localized two-dimensional grid and fit the central peak itself. This provides a joint estimate of $f$ and $\dot f$, and also makes the covariance between the two parameters explicit.

The local fitting model is motivated by the expected coherent response of a sinusoidal signal. Extending the single-frequency treatment of \citet{2023arXiv231106620S}, the expected $Z_1^2$ response can be written, up to an overall amplitude factor, as
\begin{align}
    \left\langle Z_1^2(\Delta f,\Delta \dot f)\right\rangle
    \simeq
    A
    \left|
    \frac{1}{T}
    \int_0^T
    \exp\left[
    2\pi i
    \left(
    \Delta f\,t
    +
    \frac{1}{2}\Delta \dot f\,t^2
    \right)
    \right]
    dt
    \right|^2 ,
\end{align}
where $\Delta f$ and $\Delta \dot f$ are offsets from the true rotational parameters and $A$ is the coherent peak amplitude. For the ideal sinusoidal rate model, $A$ scales approximately as $Tb^2/(2a)$.

When $\Delta \dot f=0$, this expression reduces to the usual $\mathrm{sinc}^2$ response in frequency, with characteristic width
\begin{align}
    W_f = \frac{1}{\pi T}.
\end{align}
For non-zero $\Delta \dot f$, the full expression can be written in terms of Fresnel functions, but the local behaviour near the maximum is more useful for fitting. Expanding the response to second order in start-epoch coordinates gives
\begin{align}
    \left\langle Z_1^2\right\rangle
    \simeq
    A
    \left[
    1
    -
    \frac{\pi^2T^2}{3}(\Delta f)^2
    -
    \frac{\pi^2T^3}{3}\Delta f\,\Delta\dot f
    -
    \frac{4\pi^2T^4}{45}(\Delta\dot f)^2
    \right].
\end{align}
The cross-term shows that the coherent peak is tilted in the $(f,\dot f)$ plane when the frequency is quoted at the start of the observation.

This coupling is removed, to the same order, by using the midpoint-frequency coordinate,
\begin{align}
    f_{\rm mid}
    =
    f
    +
    \frac{1}{2}\dot f T,
    \qquad
    \Delta f_{\rm mid}
    =
    \Delta f
    +
    \frac{1}{2}\Delta\dot f T .
\end{align}
In these coordinates, the local expansion becomes
\begin{align}
    \left\langle Z_1^2\right\rangle
    \simeq
    A
    \left[
    1
    -
    \frac{\pi^2T^2}{3}(\Delta f_{\rm mid})^2
    -
    \frac{\pi^2T^4}{180}(\Delta\dot f)^2
    \right],
\end{align}
with no cross-term at this order. This motivates fitting the central lobe in $(f_{\rm mid},\dot f)$ rather than in the start-epoch coordinates.

We therefore model the local two-dimensional peak as
\begin{align}
    Z_1^2(f_{\rm mid},\dot f)
    \approx
    A\,
    \mathrm{sinc}^2
    \left(
    \frac{f_{\rm mid}-f_{{\rm mid},r}}{W_f}
    \right)
    \mathrm{sinc}^2
    \left(
    \frac{\dot f-\dot f_r}{W_A}
    \right),
    \label{eq:method3_local_2d_model}
\end{align}

where $f_{{\rm mid},r}$ and $\dot f_r$ are the recovered midpoint frequency and frequency derivative. The width in the frequency-derivative direction is
\begin{align}
    W_A
    =
    \frac{\sqrt{60}}{\pi T^2}
    =
    \sqrt{15}\,W_{\mathrm{fd}},
    \qquad
    W_{\mathrm{fd}}
    =
    \frac{2}{\pi T^2}.
\end{align}
This is the same broader width scale that appears in the amplitude envelope of the coherent derivative scan.

The same local expansion gives the uncertainty estimates
\begin{align}
    \sigma(f_{{\rm mid},r})
    \simeq
    \frac{\sqrt{3}\,W_f}{\sqrt{A}},
    \qquad
    \sigma(\dot f_r)
    \simeq
    \frac{\sqrt{3}\,W_A}{\sqrt{A}} .
\end{align}
The midpoint coordinate also explains why the start-epoch frequency has a larger scatter. Since
\begin{align}
    f_{{\rm start},r}
    =
    f_{{\rm mid},r}
    -
    \frac{1}{2}\dot f_r T ,
\end{align}
and the Fisher matrix is diagonal in the local midpoint coordinates, the propagated variance is
\begin{align}
    \sigma^2(f_{{\rm start},r})
    \simeq
    \sigma^2(f_{{\rm mid},r})
    +
    \frac{T^2}{4}\sigma^2(\dot f_r)
    =
    16\,\sigma^2(f_{{\rm mid},r}) .
\end{align}
Thus,
\begin{align}
    \sigma(f_{{\rm start},r})
    \simeq
    4\,\sigma(f_{{\rm mid},r}) .
\end{align}
This factor is a consequence of the reference-epoch choice, not a physical change in the source precision.

\begin{figure*}
    \centering
    \includegraphics[width=\textwidth]{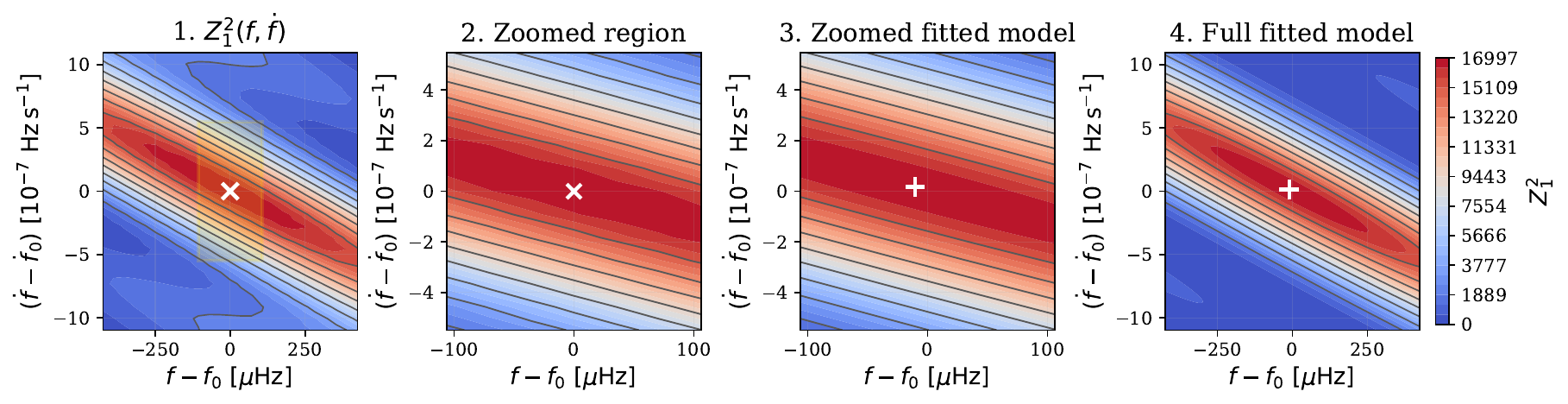}
    \caption{
Localized two-dimensional coherent fitting of a simulated $Z_1^2(f,\dot f)$ surface using the local model in Eq.~\eqref{eq:method3_local_2d_model}.
The simulation uses $a=40$, $b=30$, $f_0=5~{\rm Hz}$, $\dot f_0=-2\times10^{-5}~{\rm Hz~s^{-1}}$, and $T=1500~{\rm s}$.
The first panel shows the raw $Z_1^2(f,\dot f)$ surface evaluated on a coarse grid around the injected parameters, with the shaded rectangle marking the region selected for the local fit.
The second panel shows this selected region in detail, and the third panel shows the best-fitting local two-dimensional coherent model evaluated over the same zoomed region.
The fourth panel shows the same fitted model over the full search range.
All panels use the same colour scale; the second and third panels also use identical axis limits to allow direct comparison between the  peak and the fitted model.
The cross marks the injected parameter value, and the plus sign marks the fitted peak.
}
    \label{fig:method3_2d_fit}
\end{figure*}

\begin{algorithm}
\caption{Localized two-dimensional coherent fitting}
\label{alg:localized_2d_search}
\begin{algorithmic}[1]
\State Define a localized grid in $(f,\dot f)$ around preliminary parameter estimates.
\For{each grid point $(f,\dot f)$}
    \State Compute photon phases using the coherent phase model.
    \State Evaluate $Z_1^2(f,\dot f)$ using the full event list.
\EndFor
\State Transform the grid to midpoint coordinates using $f_{\rm mid}=f+\dot f T/2$.
\State Select the central lobe around the maximum.
\State Fit the local two-dimensional model in Eq.~\eqref{eq:method3_local_2d_model} to the selected region.
\State Record $f_{{\rm mid},r}$, $\dot f_r$, the peak amplitude $A$, and the fitted widths $W_f$ and $W_A$.
\State Transform to $f_{{\rm start},r}=f_{{\rm mid},r}-\dot f_r T/2$ if a start-epoch frequency is required.
\State Estimate $\sigma(f_{{\rm mid},r})$, $\sigma(\dot f_r)$, and $\sigma(f_{{\rm start},r})$ using the local peak-width relations.
\end{algorithmic}
\end{algorithm}

The localized two-dimensional fit is the most direct of the three approaches because it estimates the rotational parameters jointly from the coherent peak. Its cost is higher than the one-dimensional methods, but the grid can be kept local once the approximate parameter region is known. In the following section, we test these uncertainty estimates against Monte Carlo simulations and compare the scatter in $f_{{\rm start},r}$, $f_{{\rm mid},r}$, and $\dot f_r$.

\section{Monte Carlo Validation}\label{sec:validation}

The methods described above estimate rotational parameters from a single Poisson realisation of an event list. To test whether the associated uncertainty estimates have the intended frequentist meaning, we repeated the full simulation and recovery procedure over independent Poisson realisations with the same injected source parameters. Unless stated otherwise, each validation experiment uses $N_{\rm MC}=300$ realisations. This is sufficient for the comparisons made here, where the goal is to test the scaling and approximate calibration of the uncertainty estimates.

\subsection{Simulation ensemble and convergence check}\label{sec:mc_ensemble}

As a sanity check on the chosen ensemble size, we first examined how the measured scatter of recovered parameters changes as the number of Monte Carlo realisations is increased. For a fixed simulated source, we computed the recovered parameters after each independent realisation and measured the running scatter as a function of the number of realisations included. We monitored both the standard deviation and robust scatter estimators, including the median absolute deviation from the injected value.

For a recovered parameter $\theta_r$ with injected value $\theta_0$, we define
\begin{align}
    {\rm MAD}_\theta
    =
    {\rm median}\left(|\theta_r-\theta_0|\right).
\end{align}
Unlike the standard deviation, this quantity is less sensitive to occasional failed fits or secondary-lobe selections. We also compute a three-sigma-clipped standard deviation as a diagnostic check. These robust estimators are not used to redefine the uncertainty prescription; they are used only to verify that the measured scatter is not dominated by a small number of outliers.

\begin{figure}
    \centering
    \includegraphics[width=0.7\columnwidth]{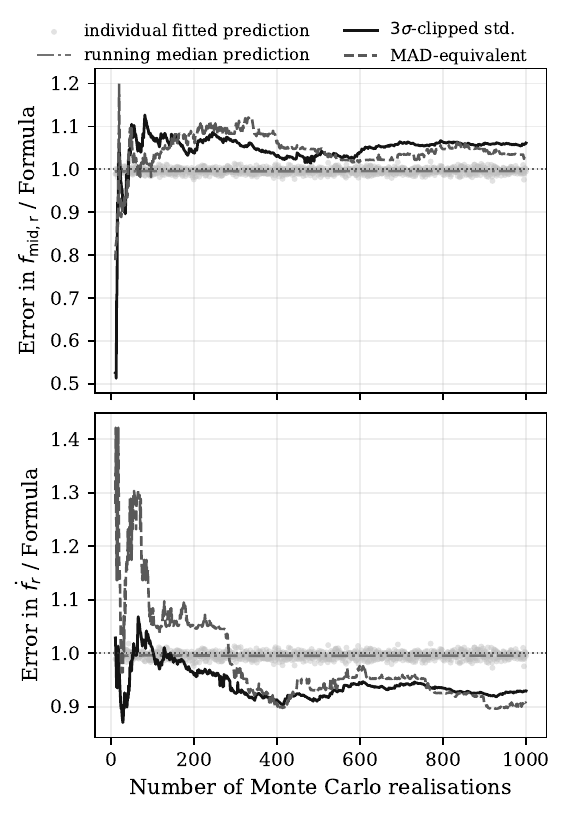}
    \caption{
    Convergence of the Monte Carlo scatter estimates as a function of the number of independent Poisson realizations included in the ensemble.
    The example shown uses the coherent derivative scan for a fixed simulated source.
    Curves show the running standard deviation, the three-sigma-clipped standard deviation, and the median absolute deviation from the injected parameter value for representative recovered parameters.
    For $N_{\rm MC}=300$, the expected fractional sampling uncertainty of a standard deviation is approximately $[2(N_{\rm MC}-1)]^{-1/2}\simeq 4$ per cent, which is adequate for the validation tests presented here.
    }
    \label{fig:mc_convergence}
\end{figure}

Figure~\ref{fig:mc_convergence} shows that the measured scatter stabilises well before the end of the ensemble for the representative cases tested. We therefore use $N_{\rm MC}=300$ as the default ensemble size in the following validation experiments.

\subsection{Validation of segmented frequency regression}
\label{sec:validation_segmented_regression}

The segmented frequency-regression method relies on estimating a local frequency from several shorter intervals and fitting those frequencies as a function of time. Its performance is therefore controlled mainly by two choices: the duration of each segment, $t_{\rm int}$, and the number of usable segments, $N_{\rm seg}$. We tested both effects using Monte Carlo realisations of the same injected signal.

\begin{figure}
    \centering
    \includegraphics[width=0.8\columnwidth]{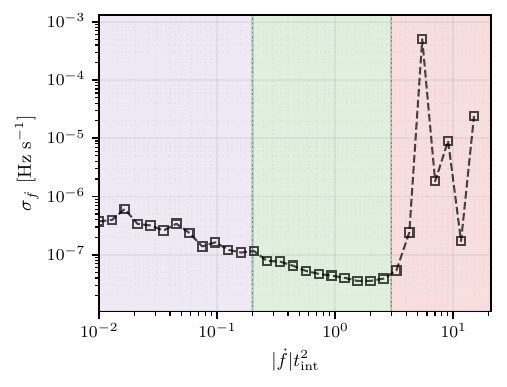}
    \caption{
    Monte Carlo validation of the segment-duration choice in segmented frequency regression.
    The recovered spin-down scatter, $\sigma(\dot f_r)$, is shown as a function of the dimensionless intra-segment phase-drift parameter $|\dot f|t_{\rm int}^2$.
    Very short segments are photon-noise limited because each local $Z_1^2(f)$ peak is weak.
    Very long segments are affected by intra-segment phase evolution, which violates the locally monochromatic approximation used to estimate the segment frequency.
    The broad intermediate region gives the practical operating range used for the segmented regression method.
    }
    \label{fig:method1_tint_validation}
\end{figure}

The segment duration sets a bias--variance trade-off. If $t_{\rm int}$ is too short, each interval contains too few photons and the local $Z_1^2(f)$ peak is weak and noisy. If $t_{\rm int}$ is too long, the frequency changes appreciably within a single interval and the assumption of a locally monochromatic signal begins to fail. The relevant dimensionless scale is $|\dot f|t_{\rm int}^2$, which measures the accumulated spin-down contribution within one segment. Figure~\ref{fig:method1_tint_validation} shows that the spin-down error is not sensitive to a finely tuned choice of $t_{\rm int}$. Rather, the error remains near its minimum over a broad intermediate range, approximately corresponding to the regime
$0.2 \lesssim |\dot f|t_{\rm int}^2 \lesssim 3$
discussed in Eq.~\ref{eq:sweet}. 

We next tested how the final parameter scatter depends on the number of usable segments. For a fixed segment duration, we repeated the segmented recovery while varying $N_{\rm seg}$ and measured the Monte Carlo scatter in the recovered frequency and frequency derivative.

\begin{figure}
    \centering
    \includegraphics[width=0.7\columnwidth]{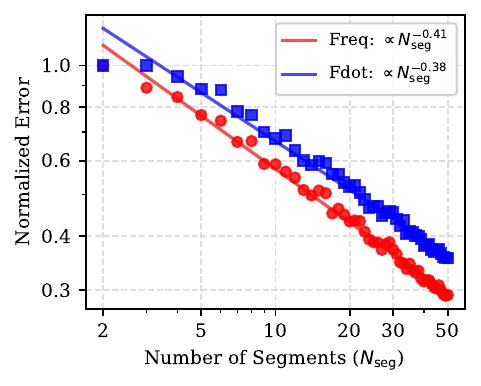}
    \caption{
    Dependence of the segmented frequency-regression accuracy on the number of usable intervals, $N_{\rm seg}$.
    The plotted values are normalized to the scatter measured at the smallest number of segments.
    The recovered frequency and frequency derivative both improve as more independent local frequency measurements are included in the weighted regression.
    The empirical scalings are shallower than the ideal $N_{\rm seg}^{-1/2}$ behaviour, reflecting the combined effects of local peak fitting, finite segment duration, and the time leverage of the sampled intervals.
    }
    \label{fig:method1_nseg_validation}
\end{figure}

Figure~\ref{fig:method1_nseg_validation} shows that the uncertainty decreases gradually as additional segments are included. The dependence is not as steep as a simple independent-measurement average because the spin-down estimate also depends on the temporal placement of the segments and on the quality of each local frequency recovery. Nevertheless, the trend demonstrates that the method is not dominated by a single interval once several usable segments are available. This makes the approach naturally suited to observations with moderate GTI losses, where empty or poorly sampled intervals can be excluded from the regression.

Together, Figs.~\ref{fig:method1_tint_validation} and \ref{fig:method1_nseg_validation} define the operational regime of the segmented method. It is most useful as a computationally inexpensive and gap-tolerant estimator of the frequency derivative, or as a preliminary step for defining the narrower coherent searches used by the methods below. Its limitation is that it does not use the full coherent power of the complete event list, and therefore in the subsequent subsection, the coherent derivative scan and local two-dimensional fit are expected to provide tighter constraints when the signal remains phase coherent over the full observation.

\subsection{Validation of the coherent estimators}
\label{sec:coherent_validation}

The convergence of the Monte Carlo scatter has already been checked in Section~\ref{sec:mc_ensemble}. We therefore do not repeat that test here. In this subsection, we focus on the two coherent estimators introduced above: the coherent derivative scan and the localized two-dimensional coherent fit. The segmented regression method provides a useful baseline, especially for interrupted observations, but it uses only locally coherent intervals. The coherent methods instead use the full event list and are therefore expected to provide tighter constraints whenever the phase model remains valid over the analysed observing span.

\begin{figure*}
    \centering
    \includegraphics[width=0.70\textwidth]{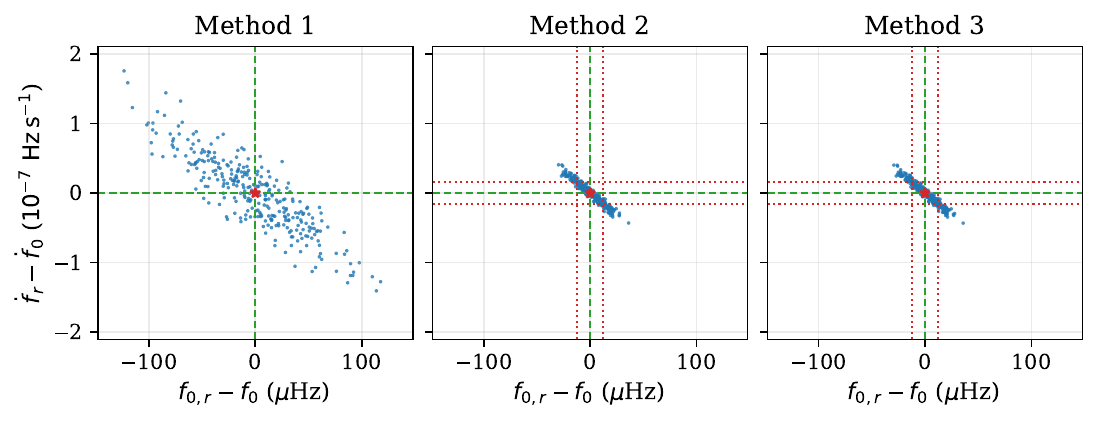}

    \vspace{0.6em}

    \includegraphics[width=0.70\textwidth]{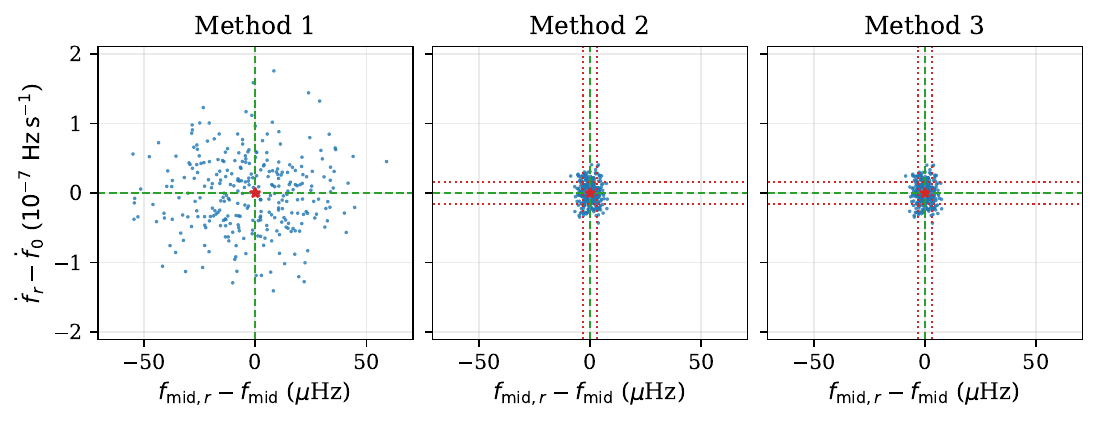}
    \caption{
    Monte Carlo scatter of the recovered rotational parameters for the three estimators.
    Top: recovery in the start-epoch frequency $f_{\rm start}$ and frequency derivative $\dot f$.
    Bottom: the same realisations expressed in terms of the midpoint frequency $f_{\rm mid}$ and $\dot f$.
    The coherent estimators show substantially smaller scatter than the segmented regression baseline.
    The midpoint-frequency parametrization also reduces the covariance between the recovered frequency and frequency derivative.
    }
    \label{fig:method123_scatter}
\end{figure*}

\begin{figure*}
    \centering
    \includegraphics[width=0.70\textwidth]{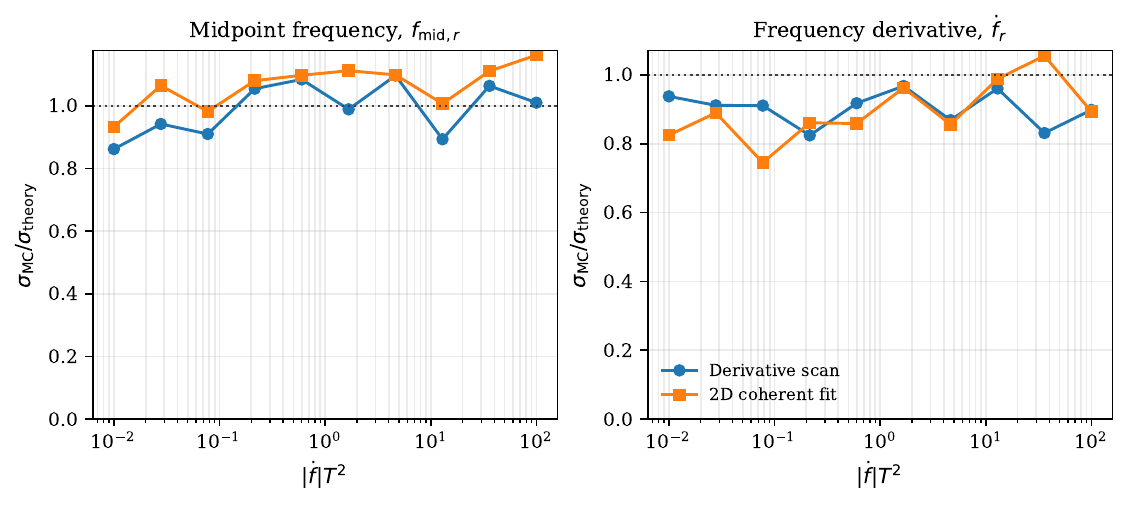}
    \caption{
    Dependence of the coherent uncertainty calibration on the total observing span $T$.
    The panels compare the Monte Carlo scatter of $f_{{\rm mid},r}$ and $\dot f_r$ with the uncertainty predicted from the fitted local peak amplitude and widths.
    For fixed source parameters and uninterrupted observations, the expected scalings are approximately $\sigma(f_{{\rm mid},r})\propto T^{-3/2}$ and $\sigma(\dot f_r)\propto T^{-5/2}$.
    }
    \label{fig:coherent_time_scaling}
\end{figure*}

\begin{figure*}
    \centering
    \includegraphics[width=0.70\textwidth]{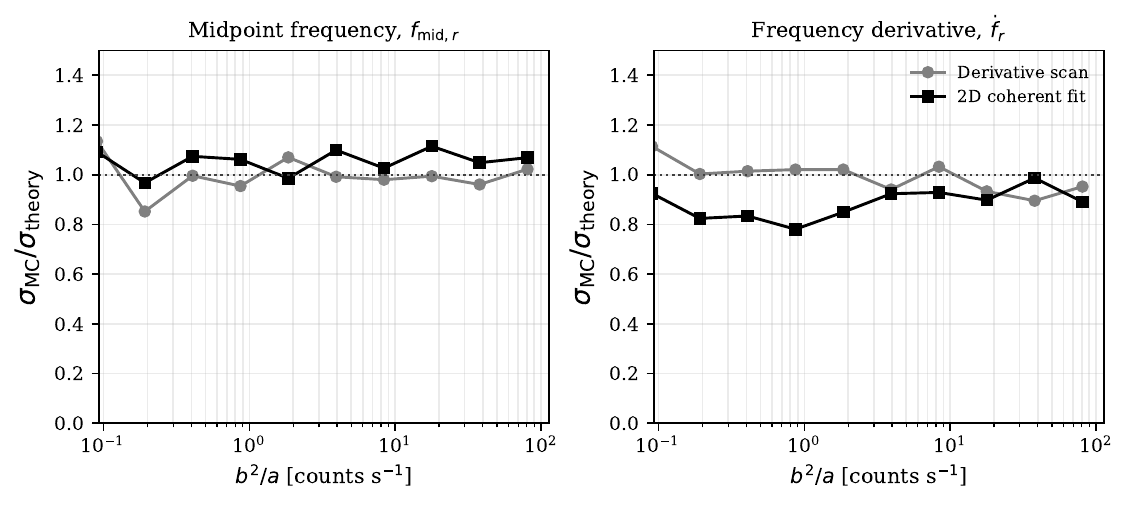}
    \caption{
    Uncertainty calibration of the two coherent estimators as a function of signal strength.
    The horizontal axis shows $b^2/a$, which sets the expected coherent peak power for fixed observing span.
    The vertical axis shows the ratio between the Monte Carlo scatter of the recovered parameter and the fitted uncertainty prediction.
    Values close to unity indicate that the fitted peak quantities provide a calibrated estimate of the statistical error.
    }
    \label{fig:coherent_strength_scaling}
\end{figure*}

For each Monte Carlo realisation, we record the recovered start-epoch frequency $f_{{\rm start},r}$, the midpoint frequency $f_{{\rm mid},r}$, and the recovered frequency derivative $\dot f_r$. The midpoint frequency is defined as
\begin{align}
    f_{{\rm mid},r}
    =
    f_{{\rm start},r}
    +
    \frac{1}{2}\dot f_r T .
\end{align}
As discussed in Section~\ref{sec:localized_2d_search}, this coordinate reduces the covariance between the recovered frequency and frequency derivative. Figure~\ref{fig:method123_scatter} shows this directly. In the start-epoch coordinates, the recovered frequency and frequency derivative are visibly correlated, because an error in $\dot f_r$ propagates into the frequency quoted at the beginning of the observation. This correlation is strongly reduced when the same realisations are expressed in terms of $f_{{\rm mid},r}$.

To quantify the calibration of the coherent uncertainty estimates, we compare the Monte Carlo scatter of the recovered parameters with the uncertainty predicted from the fitted local peak properties. For each realisation, the fitted amplitude and widths give
\begin{align}
    \widehat{\sigma}(f_{{\rm mid},r})
    &=
    \frac{\sqrt{3}\,W_{f,\rm fit}}{\sqrt{A_{\rm fit}}},
    \\
    \widehat{\sigma}(\dot f_r)
    &=
    \frac{\sqrt{3}\,W_{A,\rm fit}}{\sqrt{A_{\rm fit}}}.
\end{align}
For a given simulation ensemble, we use the calibration ratio
\begin{align}
    R_\theta
    =
    \frac{\sigma_{\rm MC}(\theta_r)}
    {{\rm median}\,[\widehat{\sigma}(\theta_r)]}.
\end{align}
Values close to unity indicate that the fitted local peak gives a reliable estimate of the run-to-run statistical scatter. We use the median predicted uncertainty because each Poisson realisation has a slightly different fitted peak amplitude and width.

We first test the expected dependence on the total observing span. For an uninterrupted observation, the local widths scale as $W_f\propto T^{-1}$ and $W_A\propto T^{-2}$. Since the coherent amplitude increases approximately linearly with $T$ for fixed source parameters, the predicted uncertainties scale approximately as
\begin{align}
    \widehat{\sigma}(f_{{\rm mid},r})
    &\propto
    T^{-3/2},
    \\
    \widehat{\sigma}(\dot f_r)
    &\propto
    T^{-5/2}.
\end{align}
Figure~\ref{fig:coherent_time_scaling} shows that the Monte Carlo scatter follows these trends for both coherent estimators. This confirms that the fitted-width prescription captures not only the shape of an individual $Z_1^2$ peak, but also the expected improvement with increasing coherent baseline.

We next vary the signal strength while keeping the search procedure fixed. For the sinusoidal rate model used here, the expected coherent peak amplitude scales approximately as
\begin{align}
    A
    \sim
    \frac{T b^2}{2a}.
\end{align}
Thus, at fixed $T$, the uncertainty is expected to decrease approximately as $(b^2/a)^{-1/2}$. Figure~\ref{fig:coherent_strength_scaling} shows the ratio of the Monte Carlo scatter to the fitted uncertainty prediction as a function of $b^2/a$. Both coherent estimators remain close to unity over the tested range, indicating that the fitted amplitude correctly captures the signal-strength dependence of the statistical error.

\begin{figure}
    \centering
    \includegraphics[width=0.82\columnwidth]{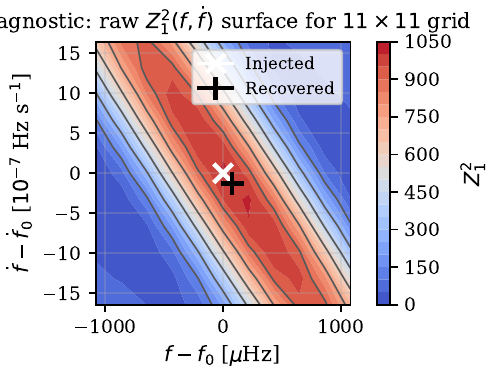}
    \caption{
    Representative raw $Z_1^2(f,\dot f)$ surface evaluated on an $11\times 11$ grid.
    The white marker denotes the injected parameters and the black marker denotes the recovered parameters after fitting the local coherent response.
    The figure illustrates the role of the raw grid in Method~3: it is used to identify and sample the coherent ridge, while the final parameter estimate is obtained from the local two-dimensional fit.
    }
    \label{fig:method3_raw_grid_example}
\end{figure}

\begin{figure}
    \centering
    \includegraphics[width=0.7\columnwidth]{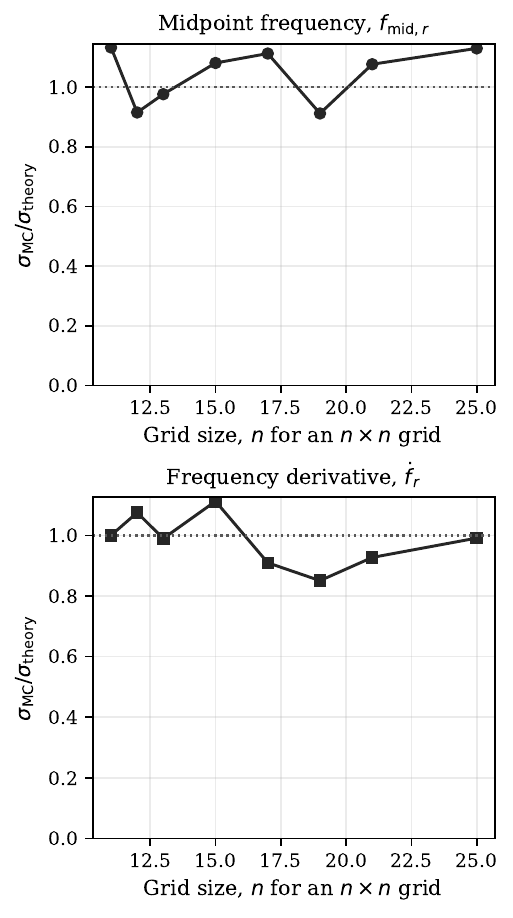}
    \caption{
    Dependence of the Method~3 uncertainty calibration on the raw two-dimensional grid resolution.
    The horizontal axis gives the grid size $n$ for an $n\times n$ search in $(f_{\rm mid},\dot f)$.
    The vertical axis shows the ratio of the Monte Carlo scatter to the fitted uncertainty prediction.
    The recovery remains stable once the local coherent peak is adequately sampled.
    }
    \label{fig:method3_grid_resolution}
\end{figure}

For Method~3, we also test how the recovery depends on the raw two-dimensional grid resolution. This is important because the method is intended to avoid an expensive dense search: the grid only needs to sample the local coherent peak well enough for the subsequent analytic fit. Figure~\ref{fig:method3_raw_grid_example} shows a representative $11\times 11$ diagnostic grid. Even at this modest resolution, the injected solution lies on the coherent ridge, and the fitted maximum is recovered from the local peak structure rather than from the grid point alone. Figure~\ref{fig:method3_grid_resolution} quantifies this behaviour by repeating the recovery for different raw grid resolutions. The uncertainty ratios stabilize once the local peak is sampled by a moderate number of grid points. Thus, Method~3 does not require an exhaustive dense search over the full two-dimensional parameter space; a coarse local grid followed by a fitted coherent response is sufficient in the tested regime.

\begin{figure}
    \centering
    \includegraphics[width=0.92\columnwidth]{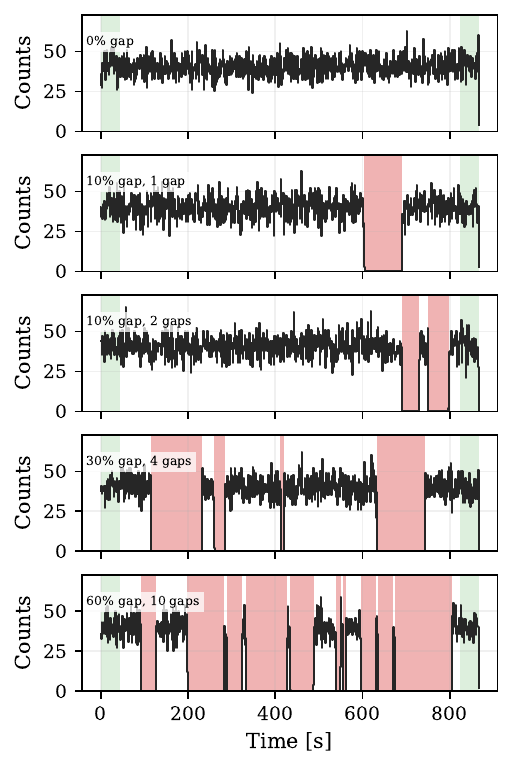}
    \caption{
    Representative good-time-interval patterns used in the gap-robustness tests.
    The black curves show binned event counts, the red shaded regions mark removed observing intervals, and the green shaded regions mark the protected beginning and end of the observation.
    The examples illustrate that a given removed observing fraction can be distributed in qualitatively different ways across the observation, thereby changing the temporal window function in addition to reducing the number of photons.
    }
    \label{fig:gti_gap_patterns}
\end{figure}

\begin{figure}
    \centering
\includegraphics[width=0.70\columnwidth]{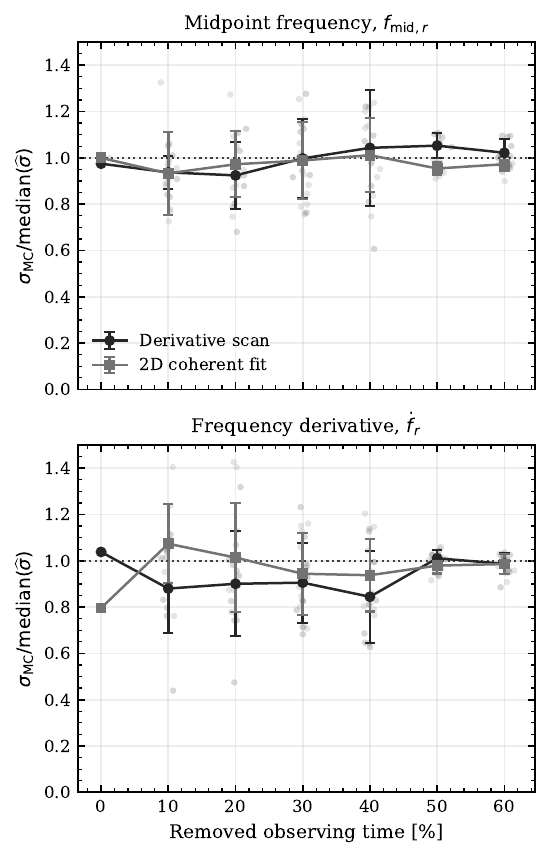}
    \caption{
    Robustness of the coherent uncertainty estimates to observational gaps.
    The horizontal axis shows the fraction of removed observing time.
    For each gap fraction, several random gap patterns are tested.
    The vertical axis shows the ratio between the Monte Carlo scatter of the recovered parameter and the median fitted uncertainty prediction from the individual realisations.
    The dotted horizontal line marks perfect calibration.
    Both coherent estimators remain close to unity for the midpoint frequency and frequency derivative, showing that fitting the local coherent response provides a practical way to account for the modified window function.
    }
    \label{fig:gti_gap_error_ratios}
\end{figure}

Finally, we test robustness to observational gaps. For each removed-time fraction, we generate artificial good-time-interval patterns by distributing the removed observing time into one or more internal gaps, while keeping the beginning and end of the observation protected. This construction is useful because it separates two effects: the loss of photons and the modification of the temporal window function. Figure~\ref{fig:gti_gap_patterns} shows representative examples. The same removed observing fraction can correspond to a single long interruption or to several shorter interruptions, and these cases need not affect the coherent peak in the same way.

For the gapped simulations, we do not use the ideal uninterrupted-window expressions for $W_f$ and $W_A$. Instead, the uncertainty prediction for each realisation is computed from the fitted amplitude and fitted widths of that same gapped event list. This choice is important because gaps alter the local shape and width of the coherent response. Figure~\ref{fig:gti_gap_error_ratios} shows that the coherent uncertainty estimates remain close to the Monte Carlo scatter even when a substantial fraction of the observing time is removed. The scatter between different gap patterns increases at large removed fractions, as expected, because the window function can distort the coherent peak in different ways. Nevertheless, the fitted local response absorbs much of this effect, provided that the central peak remains identifiable and well sampled.

Taken together, these tests show that the coherent estimators provide calibrated uncertainty estimates over the regimes tested here. Method~2 gives an interpretable one-dimensional refinement of the frequency derivative through the amplitude response envelope, while Method~3 gives a compact joint fit in the $(f_{\rm mid},\dot f)$ plane. The midpoint-frequency coordinate is especially useful because it separates the two fitted parameters to leading order and gives a more stable frequency estimate than the start-epoch coordinate. In the following section, we apply these calibrated estimators to the observed pulsar event lists.

\section{Application to AstroSat event lists}
\label{sec:results}

\begin{figure*}
    \centering
    \includegraphics[width=0.82\textwidth]{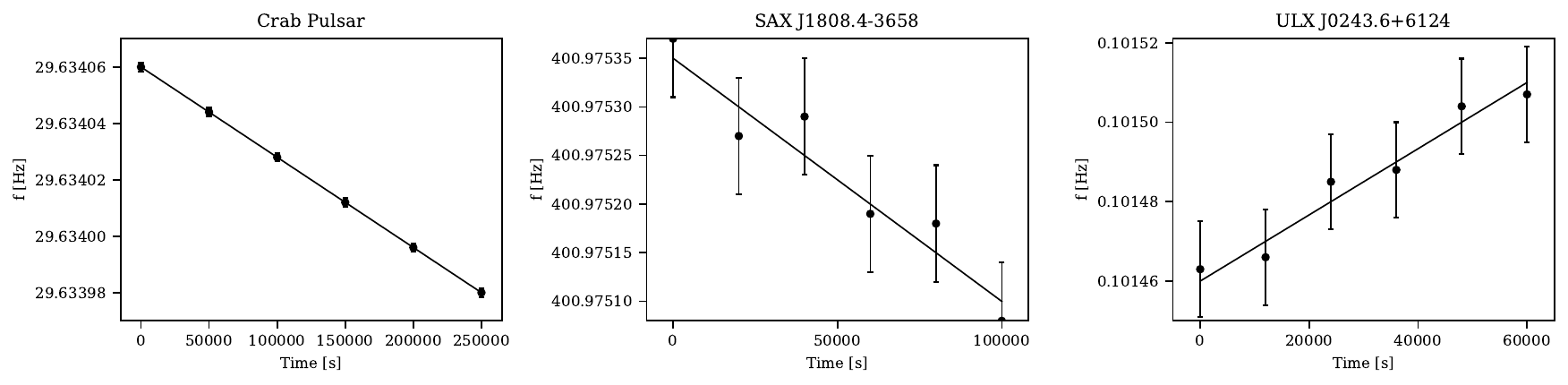}
    \caption{
Segmented frequency recovery for the three AstroSat event lists using Method~I.
Left: Crab pulsar, analysed using six independent intervals of duration $20000$~s from a longer event list of duration $\sim 300000$~s.
Middle: SAX~J1808.4$-$3658, analysed using intervals of duration $10000$~s from an event list of duration $\sim 100000$~s.
Right: Swift~J0243.6+6124, analysed using intervals of duration $1000$~s from an event list of duration $\sim 70000$~s.
The points show the locally recovered frequencies from one-dimensional $Z_1^2(f)$ searches, and the straight lines show weighted linear fits used to estimate $\dot f$.
}
    \label{fig:all_method1}
\end{figure*}

\begin{figure*}
    \centering
    \includegraphics[width=0.78\textwidth]{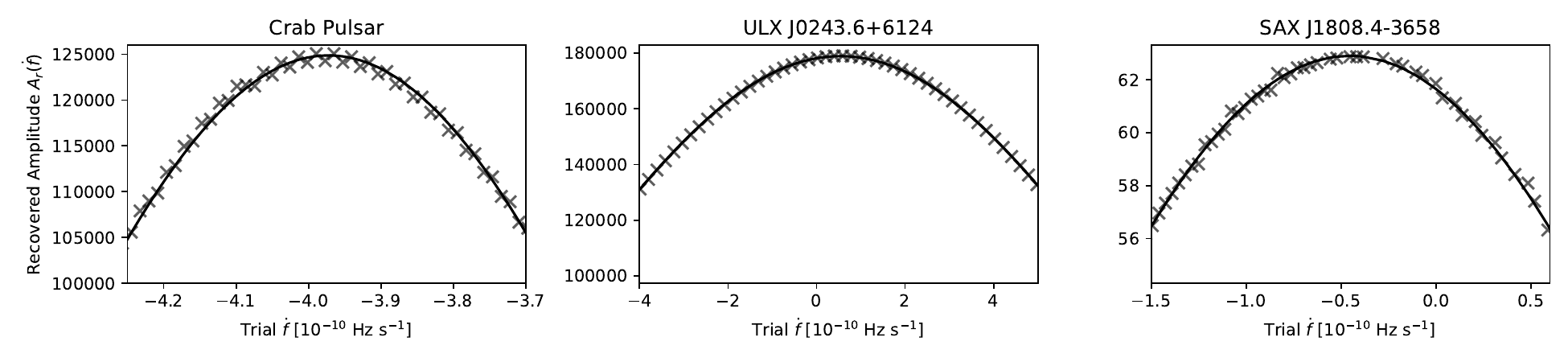}
    \caption{
    Coherent derivative scan for the Crab pulsar, Swift~J0243.6+6124, and SAX~J1808.4$-$3658 using Method~II.
    For each trial frequency derivative $\dot f_{\rm trial}$, a local one-dimensional $Z_1^2(f)$ peak is fitted and its recovered amplitude $A_r$ is recorded.
    The resulting amplitude envelope is then fitted with a local $\mathrm{sinc}^2$ response to estimate the preferred $\dot f$.
    The fitted peak location gives the recovered frequency derivative, while the fitted amplitude and width provide the corresponding uncertainty estimate.
    }
   \label{fig:method2_all_sources}
\end{figure*}

\begin{figure*}
    \centering
    \includegraphics[width=0.82\textwidth]{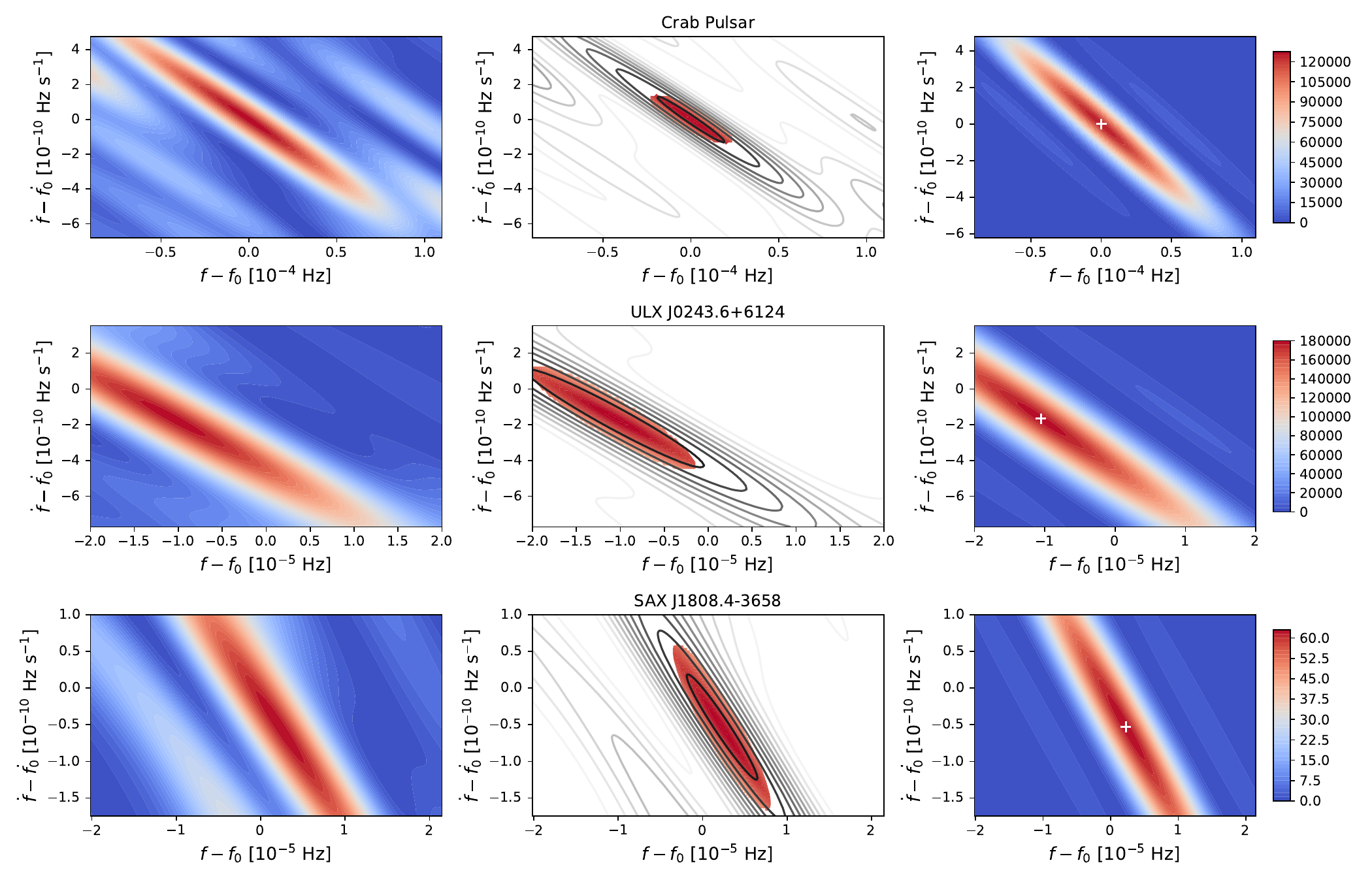}
    \caption{
    Two-dimensional coherent recovery for the Crab pulsar, Swift~J0243.6+6124, and SAX~J1808.4$-$3658 using Method~III.
    Each row corresponds to one source.
    The left panels show the raw two-dimensional $Z_1^2$ surface in the local $(f,\dot f)$ search region.
    The middle panels show the local region used for fitting the central coherent response.
    The right panels show the corresponding fitted two-dimensional response model.
    The white markers indicate the recovered rotational parameters.
    All axes are plotted relative to the adopted reference frequency and frequency derivative for each source.
    }
   \label{fig:method3_all_sources}
\end{figure*}

Having calibrated the three estimators on controlled Poisson simulations, we now apply them to AstroSat event lists for the Crab pulsar, Swift~J0243.6+6124, and SAX~J1808.4$-$3658. The goal of this section is not to reintroduce the methods, but to demonstrate how they behave on real event data with source-dependent count rates, observing spans, and good-time-interval structure. We use Method~I as a diagnostic baseline for visualising spin evolution, and Methods~II and III as coherent estimators for the rotational parameters and their uncertainties.

Figure~\ref{fig:all_method1} shows the segmented frequency tracks obtained with Method~I.
For the Crab pulsar, the long observing span allows the event list to be divided into several usable intervals, and the recovered frequencies show a clear monotonic trend.
For SAX~J1808.4$-$3658 and Swift~J0243.6+6124, the available spans and source properties make the segmentation less favourable.
In these cases, the method still provides a useful visual check on the approximate spin evolution, but the inferred derivative is more sensitive to the chosen interval duration and to the number of usable segments.
We therefore treat Method~I primarily as a diagnostic and initialization tool, rather than as the final estimator for the real-data analysis.

We next apply the coherent derivative scan, Method~II. The approximate derivative range is guided by the segmented trend, but the final estimate is obtained from the coherent response of the event list. Figure~\ref{fig:method2_all_sources} shows the recovered amplitude envelope $A_r(\dot f_{\rm trial})$ for the three sources. In this method, an incorrect trial derivative reduces the coherent peak height and shifts the frequency at which the one-dimensional $Z_1^2(f)$ profile is maximized. The maximum of the amplitude envelope therefore gives a direct estimate of the preferred $\dot f$. Compared with Method~I, this procedure uses a larger coherent fraction of the data and gives a more stable estimate of the frequency derivative.

Finally, we apply the localized two-dimensional coherent fit, Method~III. This method fits the local $Z_1^2$ surface jointly in frequency and frequency derivative, rather than scanning the derivative and fitting one-dimensional peaks separately. Figure~\ref{fig:method3_all_sources} shows the raw two-dimensional search region, the local fitting region, and the fitted coherent response for the three sources. In practice, the preceding methods provide a useful guide to the location and scale of the local search region, so that Method~III can be applied without an exhaustive dense search over a large parameter space. The fitted two-dimensional response gives both the recovered rotational parameters and the uncertainty estimates through the fitted amplitude and widths.

The coherent methods give mutually consistent derivative estimates within the expected level of scatter for the three sources, while Method~III also provides a direct estimate of the frequency at the adopted reference epoch. The results are summarized in Table~\ref{tab:combined_methods}. The comparison should be interpreted with the different roles of the methods in mind: Method~I is useful for visualizing the secular trend and detecting gross timing behaviour, Method~II gives a coherent derivative estimate through the amplitude envelope, and Method~III gives the most compact joint recovery of frequency and frequency derivative.
\begin{table}
\centering
\caption{
Recovered rotational parameters for the three AstroSat event lists using the three estimators. 
The quoted frequency is evaluated at the adopted reference epoch used for each source. 
The frequency derivative is reported in units of $10^{-10}\,{\rm Hz\,s^{-1}}$, with the sign shown explicitly for spin-up cases. 
Uncertainties for the coherent methods are obtained from the fitted local peak amplitudes and widths, as calibrated in Section~\ref{sec:validation}.}

\label{tab:combined_methods}
\begin{tabular}{
    l
    S[table-format=3.4]
    @{\,${}\pm{}$\,}
    S[scientific-notation=true, table-format=1.1e-1]
    S[table-format=+2.3]
    @{\,${}\pm{}$\,}
    S[table-format=1.3]
}
\hline
{Source / Method}
& \multicolumn{2}{c}{$f_{\rm{start}}$ (Hz)}
& \multicolumn{2}{c}{$\dot{f}$ ($10^{-10}$ Hz s$^{-1}$)} \\
\hline

\multicolumn{5}{l}{\textbf{Crab Pulsar}} \\ 
Method I
& 29.6340 & 1.1e-7
& -3.716 & 0.007 \\
Method II
& 29.6342 & 1.2e-8
& -3.975 & 0.0018 \\
Method III
& 29.6342 & 1.6e-8
& -3.978 & 0.0011 \\

\hline
\multicolumn{5}{l}{\textbf{Swift J0243.6+6124}} \\
Method I
& 0.1014 & 1.5e-5
& +36.8 & 3.5 \\
Method II
& 0.1015 & 1.2e-8
& +0.537 & 0.019 \\
Method III
& 0.1015 & 1.3e-8
& +0.576 & 0.014 \\

\hline
\multicolumn{5}{l}{\textbf{SAX J1808.4$-$3658}} \\
Method I
& 400.975 & 2.1e-5
& -25.1 & 4.7 \\
Method II
& 400.974 & 2.5e-7
& -0.44 & 0.40 \\
Method III
& 400.975 & 4.1e-7
& -0.42 & 0.26 \\

\hline
\end{tabular}
\end{table}

\begin{figure}
    \centering
    \includegraphics[width=0.92\columnwidth]{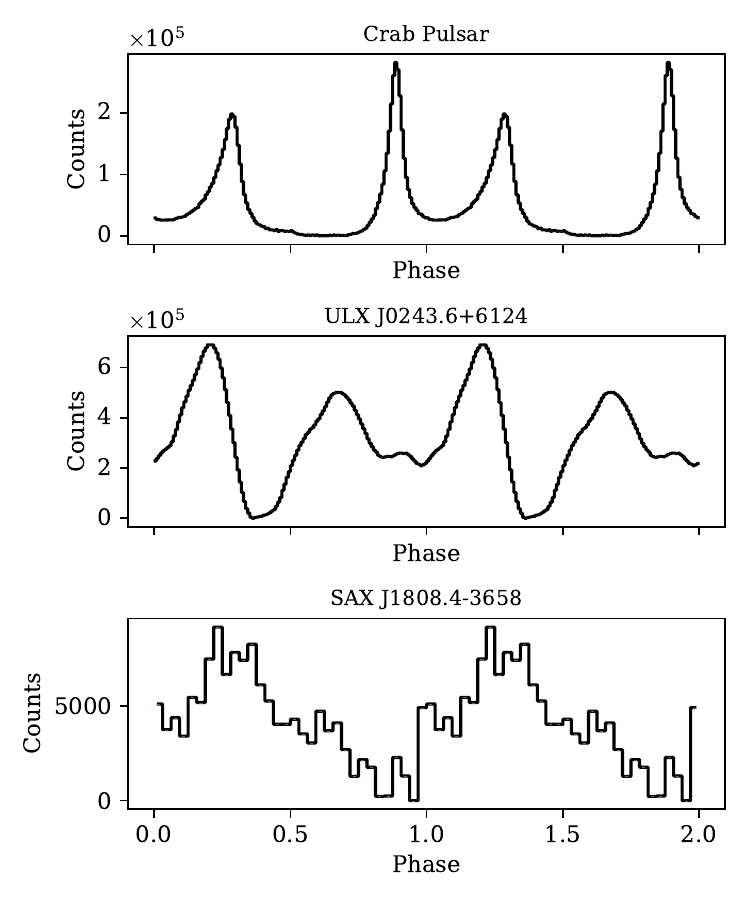}
    \caption{
Phase-folded pulse profiles obtained using the recovered rotational solutions.
The profiles are shown over two rotational cycles and normalized by the total number of photons.
From left to right, the panels show: the Crab pulsar, with its characteristic main pulse and interpulse structure; Swift~J0243.6+6124, with a broad and structured pulse profile; and SAX~J1808.4$-$3658, with the approximately sinusoidal modulation typical of this accreting millisecond X-ray pulsar.
The start epochs for the folding are MJD~58148.688 for the Crab pulsar, MJD~58033.199 for Swift~J0243.6+6124, and MJD~58709.054 for SAX~J1808.4$-$3658.
}
    \label{fig:phase_sources}
\end{figure}

As a final consistency check, we fold the event lists using the recovered rotational solutions. The resulting pulse profiles are shown in Figure~\ref{fig:phase_sources}. The Crab profile shows the expected main-pulse and interpulse structure, SAX~J1808.4$-$3658 shows an approximately sinusoidal modulation, and Swift~J0243.6+6124 shows a broader and more structured pulse profile. These folded profiles are not used to define the uncertainty estimates, but they provide a useful physical check that the recovered rotational parameters produce coherent pulsations in the observed event lists.

To check whether the uncertainty estimates derived from the local coherent response are applicable to the real event lists, we performed simplified source-matched simulations. For each source, we used the nominal rotational parameters, approximate count-rate scale, and good-time-interval structure of the corresponding AstroSat observation. These simulations are not intended to reproduce the full astrophysical complexity of the sources, including detailed pulse morphology, spectral variability, or non-stationary accretion behaviour. Rather, they provide controlled Poisson event lists with comparable time coverage, gaps, and count rates. Across repeated realizations, the scatter in the recovered rotational parameters is consistent with the predicted uncertainty scale within the expected Monte Carlo fluctuations, as illustrated for the Crab-like case in Fig.~\ref{fig:mc_convergence_crab}. Additional source-matched simulations, including the observing-window structure used for the three event lists, are shown in Appendix~\ref{app:source_matched_simulations}.

\begin{figure}
\centering
\includegraphics[
    width=0.8\columnwidth,
    trim={0cm 7cm 0cm 0cm},
    clip
]{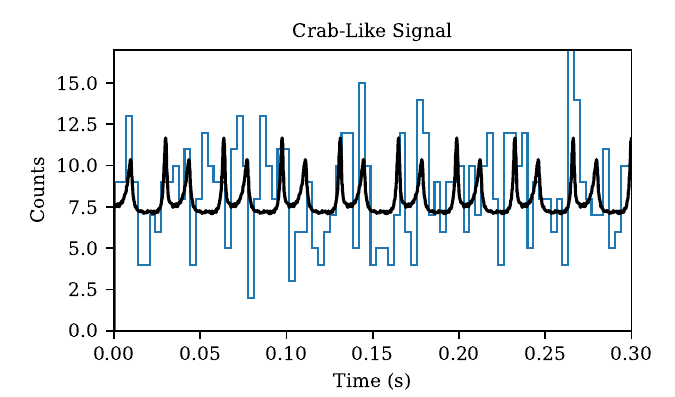}
\caption{
Representative simulated Crab-like event series used for validating the uncertainty estimates.
The injected sinusoidal signal has $(a,b,f,\dot f,T)=(2200,550,29.634065~{\rm Hz},-3.517\times10^{-10}~{\rm Hz~s^{-1}},80000~{\rm s})$, chosen to be comparable to the AstroSat/LAXPC Crab event list analysed in this work.
The photon arrival times are binned here at $0.0033~{\rm s}$ only for visualization; the parameter recovery and uncertainty estimation are performed using the event data.
}
\label{fig:ulx_crab_toa}
\end{figure}

\begin{figure}
    \centering
    \includegraphics[width=0.8\columnwidth]{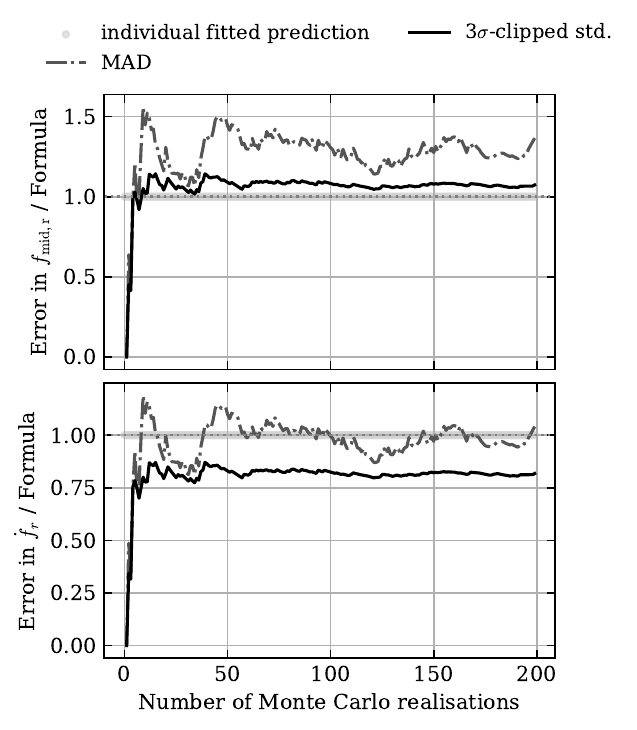}
    \caption{
    Convergence of the recovered-parameter scatter for the simulated Crab-like signal shown in Fig.~\ref{fig:ulx_crab_toa}.
    We generate $N_{\rm MC}=200$ independent Poisson realizations of the same injected sinusoidal signal and recover the rotational parameters using the localized two-dimensional coherent scan.
    The curves show the running standard deviation, the three-sigma-clipped standard deviation, and the median-absolute-deviation-based scatter estimate for representative recovered parameters.
    The convergence of these scatter estimates provides a direct check that the fitted uncertainty estimates are consistent with the frequentist spread of the recovered parameters for this source configuration.
    }
    \label{fig:mc_convergence_crab}
\end{figure}


\section{Discussion and conclusions}
\label{sec:discussion}

We have developed and tested a $Z_1^2$-based framework for recovering the spin frequency and frequency derivative of Poisson-limited high-energy pulsars. The main point of this work is that the location of the detection peak and the uncertainty on that location are separate statistical questions. The width of a $Z_1^2$ peak describes the scale over which phase coherence is lost, but the scatter of the recovered maximum also depends on the coherent signal strength and on the local geometry of the search surface. We have therefore combined analytic arguments, local peak modelling, and Monte Carlo simulations to calibrate the recovery of $f$ and $\dot f$ in the regimes considered here.

This distinction is important for follow-up work. In any timing problem, underestimated or poorly calibrated errors can propagate into subsequent modelling, source classification, braking-index estimates, torque studies, or follow-up searches. Similar concerns arise in continuous gravitational-wave searches, where parameter uncertainties determine the size of the follow-up parameter space and the computational cost of coherent refinements \citep[e.g.][]{Shaltev2014,Singhal2019,Piccinni2019}. In this sense, our results are consistent with the broader lesson from both high-energy pulsar searches and continuous-wave pipelines: efficient recovery is useful only if the uncertainty scale is also understood.

We compared three estimators. Segmented frequency regression is a useful diagnostic method: it gives a direct visual picture of spin evolution, handles gaps naturally, and can help identify a narrow region for coherent searches. Its performance, however, depends on the segment duration and on the number and placement of usable intervals. Very short segments are photon-noise limited, while very long segments suffer from intra-segment phase drift. We therefore regard it mainly as a robust exploratory tool and an initializer for coherent methods.

The coherent derivative scan improves on this by retaining the full event list while scanning trial values of $\dot f$. The amplitude envelope $A_{\rm pk}(\dot f_{\rm trial})$ provides an efficient way to refine the frequency derivative without a full dense two-dimensional search. It is also useful as a benchmark in crowded periodograms, where several candidate peaks may appear for the same source. A genuine spin-evolving signal should show a consistent evolution of peak height and recovered peak frequency as $\dot f_{\rm trial}$ is varied, whereas noise peaks, windowing artefacts, or unrelated instrumental structures are less likely to follow this behaviour. Thus, Method II is not only a refinement step, but also a useful diagnostic for assessing whether a candidate peak is compatible with the expected spin evolution of the source.

The localized two-dimensional coherent fit gives the most direct joint recovery of $f$ and $\dot f$. Expressing the problem in terms of the midpoint frequency $f_{\rm mid}$ removes the leading-order covariance between frequency and frequency derivative, and leads to simple uncertainty estimates of the form
\begin{align}
    \widehat{\sigma}(f_{{\rm mid},r})
    &=
    \frac{\sqrt{3}\,W_{f,\rm fit}}{\sqrt{A_{\rm fit}}},
    &
    \widehat{\sigma}(\dot f_r)
    &=
    \frac{\sqrt{3}\,W_{A,\rm fit}}{\sqrt{A_{\rm fit}}}.
\end{align}
Monte Carlo simulations show that these fitted-width estimates reproduce the run-to-run scatter over the tested ranges of observing span, signal strength, grid resolution, and good-time-interval structure. We therefore recommend Method III for the final joint estimate of the rotational parameters and their uncertainties, while Method II provides a computationally efficient and generally reliable intermediate estimate and consistency check.

A practical workflow follows from these results. For a targeted high-energy periodic source with an approximate known period and spin derivative, one may first use segmented regression to check whether the segment-wise frequencies show a broadly linear trend and whether gaps, glitches, or strong non-stationarity are likely to affect the analysis. This gives a preliminary region in $(f,\dot f)$. The coherent derivative scan can then be used to refine $\dot f$, reject inconsistent candidate peaks, and identify the local region for the final fit. Finally, the localized two-dimensional coherent fit in $(f_{\rm mid},\dot f)$ gives the joint parameter estimate and calibrated uncertainty. In well-behaved targeted searches, this avoids the need for an exhaustive Monte Carlo calibration for every observation, although source-matched simulations remain useful validation checks.

We have focused on targeted, narrow-band searches using the fundamental harmonic $Z_1^2$, with stable spin evolution over the analysed interval. The numerical choices for segment lengths, derivative grids, peak-selection regions, and fitting windows are chosen for this regime. Strong timing noise, unresolved glitches, binary-orbital residuals, non-stationary backgrounds, large profile changes, or highly irregular observing windows can distort the coherent peak and require manual inspection or source-specific simulations before applying the formulae directly. Thus, the method is intended as a calibrated procedure for directed searches in the tested regime, not as a fully automatic prescription for all periodic X-ray sources.

In summary, segmented regression provides a transparent diagnostic of spin evolution, the coherent derivative scan gives an efficient refinement of $\dot f$ and a useful consistency test in crowded search spaces, and the localized two-dimensional fit gives the cleanest joint recovery with calibrated uncertainties. Together, these methods provide a practical route from sparse photon arrival times to uncertainty-aware rotational-parameter estimates for high-energy pulsars and related Poisson-limited periodic sources. The next step is to incorporate this methodology into reusable timing-analysis pipelines, with additional validation for more complex pulse shapes, observing gaps, orbital residuals, and higher-order spin evolution.

\section*{Data availability}

The \textit{AstroSat}/LAXPC data used in this work are publicly available from the Indian Space Science Data Centre archive. The simulated event lists and analysis scripts used to reproduce the validation tests and figures will be shared on reasonable request to the corresponding author.

\section*{Acknowledgements}

We thank Varun Bhalerao and Utkarsh Pathak for their support in identifying and obtaining the \textit{AstroSat} source event files used in this study.
During the preparation of this manuscript, AI-assisted language tools were used to improve clarity and wording. All scientific content, analysis, derivations, figures, interpretations, and conclusions were produced, checked, and approved by the authors.

\section{Appendix}
\subsection{Alternative implementation of the two-dimensional peak fit}
\begin{figure}
    \centering
    \includegraphics[width=0.7\linewidth]{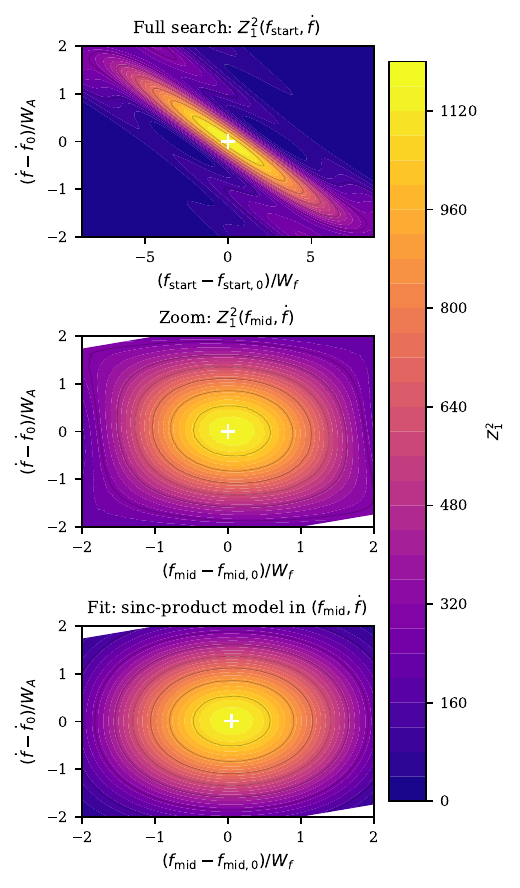}
    \caption{
    Illustration of the midpoint-frequency reparametrization for a single simulated pulsar event list.
    The events are generated using a sinusoidal count-rate model with mean rate $a=80~{\rm counts~s^{-1}}$, pulse amplitude $b=40~{\rm counts~s^{-1}}$, start-epoch frequency $f_{\rm start,0}=350~{\rm Hz}$, frequency derivative $\dot{f}_0=-2\times10^{-5}~{\rm Hz~s^{-1}}$, and exposure time $T=120~{\rm s}$.
    Top: the $Z_1^2$ search surface in the original $(f_{\rm start},\dot{f})$ coordinates, shown in units of the natural frequency width $W_f=1/(\pi T)$ and derivative width $W_{A}=\sqrt{60}/(\pi T^2)$.
    The peak appears tilted because an error in $\dot{f}$ can be partly compensated by an opposite shift in the start-epoch frequency.
    Middle: the same $Z_1^2$ values plotted after transforming to the midpoint frequency, $f_{\rm mid}=f_{\rm start}+\dot{f}T/2$, which largely removes the local correlation between frequency and frequency derivative.
    Bottom: the corresponding local sinc-product fit in $(f_{\rm mid},\dot{f})$, used to estimate the peak position.
    }
    \label{fig:z1sq_fmid_reparametrization}
\end{figure}
\label{app:curve_fit_midpoint}

The local peak-fitting procedure described in the main text can also be implemented directly as a two-dimensional curve fit to the computed $Z^2_1$ search surface. In this form, the search is first carried out on a grid in the start-epoch parameters $(f_{\rm start},\dot{f})$, and the fitted coordinates are then expressed in terms of the midpoint frequency
\begin{align}
    f_{\rm mid} = f_{\rm start} + \frac{T}{2}\dot{f},
\end{align}
where $T$ is the duration of the observation. The same computed values of $Z^2_1$ are therefore reparametrized from $(f_{\rm start},\dot{f})$ to $(f_{\rm mid},\dot{f})$ before fitting the local peak.

In practice, the fitting region may be selected in several equivalent ways. A simple rectangular window around the grid maximum is often sufficient, especially when the grid is already sampled in units of the natural widths $W_f$ and $W_{\dot f}$. Alternatively, an elliptical region in the $(f_{\rm mid},\dot{f})$ plane, or a region selected by a fixed drop in $Z^2_1$ from the maximum, can be used to avoid including points from the outer wings of the peak. We find that these choices give consistent parameter estimates, provided that the fitted region is restricted to the local central peak and does not include neighbouring side lobes or unrelated structure in the search surface.

The curve-fitting model used for this local description is
\begin{align}
    Z^2_1(f_{\rm mid},\dot{f})
    =
    A~{\rm sinc}^2\!\left(\frac{f_{\rm mid}-f_{\rm mid,0}}{W_f}\right)
    {\rm sinc}^2\!\left(\frac{\dot{f}-\dot{f}_0}{W_{A}}\right),
\end{align}
where $A$ is the fitted peak amplitude, and $(f_{\rm mid,0},\dot{f}_0)$ are the fitted peak parameters. Here
\begin{align}
    W_f = \frac{1}{\pi T},
    \qquad
    W_{A} = \frac{\sqrt{60}}{\pi T^2},
\end{align}
are the natural coherent widths in the midpoint-frequency parametrization.

This implementation was tested on single-realization simulations and repeated simulations over a representative range of source and observing parameters. These tests included different signal-to-noise ratios, photon count rates, pulse amplitudes, observing durations, trial frequencies, frequency derivatives, and sampling gaps. For parameters typical of many rotation-powered and accreting X-ray pulsars, with spin periods ranging from a few milliseconds to several seconds and with $|\dot{f}|$ spanning roughly $10^{-17}$--$10^{-5}\,{\rm Hz\,s^{-1}}$ in ordinary timing applications, the midpoint-coordinate fit recovers the injected rotational parameters within the expected statistical uncertainty. The same framework also remains applicable to accreting systems, where $\dot{f}$ may be either positive or negative depending on the instantaneous torque.

The purpose of this curve-fitting form is not to replace the full search, but to provide a compact local description of the peak once the relevant region of parameter space has been identified. It also gives a practical way to estimate the peak strength and local widths directly from the measured $Z^2_1$ surface. Within the tested regimes, the fitted midpoint frequency and frequency derivative are stable against reasonable changes in the fitting window, and the recovered uncertainties are consistent with the expected frequentist scatter of the recovered parameters.

\subsection{Extension to a search including $\ddot f$}
\label{app:fddot_extension}

The coherent response discussed in the main text can be extended formally to a search that includes a second frequency derivative. In this case the rotational phase may be written as
\begin{align}
    \psi(t)
    =
    f_{\rm start} t
    +
    \frac{1}{2}\dot f_{\rm start} t^2
    +
    \frac{1}{6}\ddot f\,t^3 ,
\end{align}
where the frequency and first derivative are quoted at the start of the observation. For an observation of duration $T$, it is useful to introduce the midpoint time coordinate
\begin{align}
    \tau = t-\frac{T}{2}.
\end{align}
The corresponding midpoint frequency and first derivative are
\begin{align}
    f_{\rm mid}
    &=
    f_{\rm start}
    +
    \frac{T}{2}\dot f_{\rm start}
    +
    \frac{T^2}{8}\ddot f ,
    \\
    \dot f_{\rm mid}
    &=
    \dot f_{\rm start}
    +
    \frac{T}{2}\ddot f .
\end{align}
However, when $\ddot f$ is included, $f_{\rm mid}$ is not the fully decorrelated frequency-like coordinate. This is because the cubic phase term is still correlated with the linear phase term over a finite observing span. Orthogonalizing the linear and cubic terms over the interval $-T/2\leq \tau \leq T/2$ gives
\begin{align}
    \tau^3
    \rightarrow
    \tau^3-\frac{3T^2}{20}\tau .
\end{align}
The natural frequency-like coordinate is therefore
\begin{align}
    F
    &=
    f_{\rm mid}
    +
    \frac{T^2}{40}\ddot f
    \\
    &=
    f_{\rm start}
    +
    \frac{T}{2}\dot f_{\rm start}
    +
    \frac{3T^2}{20}\ddot f .
\end{align}
In terms of the local parameter offsets $(\Delta F,\Delta\dot f_{\rm mid},\Delta\ddot f)$, the phase mismatch can then be written, up to an irrelevant constant phase term, as
\begin{align}
    \Delta\psi(\tau)
    =
    \Delta F\,\tau
    +
    \frac{1}{2}\Delta\dot f_{\rm mid}
    \left(\tau^2-\frac{T^2}{12}\right)
    +
    \frac{1}{6}\Delta\ddot f
    \left(\tau^3-\frac{3T^2}{20}\tau\right).
\end{align}
These three basis functions are mutually orthogonal over the observing interval. This suggests that the local coherent peak can be approximated in the decorrelated coordinates by a separable form,
\begin{align}
    Z_1^2(F,\dot f_{\rm mid},\ddot f)
    \simeq
    C
    +
    A
    \,
    {\rm sinc}^2\!\left(\frac{F-F_0}{W_f}\right)
    {\rm sinc}^2\!\left(\frac{\dot f_{\rm mid}-\dot f_0}{W_{\dot f}}\right)
    {\rm sinc}^2\!\left(\frac{\ddot f-\ddot f_0}{W_{\ddot f}}\right),
    \label{eq:fddot_local_model}
\end{align}
where $C$ is a local baseline and $A$ is the coherent peak amplitude. The corresponding natural widths are
\begin{align}
    W_f
    &=
    \frac{1}{\pi T},
    \\
    W_{\dot f}
    &=
    \frac{\sqrt{60}}{\pi T^2},
    \\
    W_{\ddot f}
    &=
    \frac{\sqrt{8400}}{\pi T^3}.
\end{align}
For a sinusoidal signal with mean count rate $a$ and pulsed amplitude $b$, the expected coherent amplitude scale is
\begin{align}
    A
    \simeq
    \frac{Tb^2}{2a}.
\end{align}
The corresponding approximate one-sigma uncertainties are therefore
\begin{align}
    \sigma_F
    &\simeq
    \frac{\sqrt{3}W_f}{\sqrt{A}},
    \\
    \sigma_{\dot f_{\rm mid}}
    &\simeq
    \frac{\sqrt{3}W_{\dot f}}{\sqrt{A}},
    \\
    \sigma_{\ddot f}
    &\simeq
    \frac{\sqrt{3}W_{\ddot f}}{\sqrt{A}}.
\end{align}
Equivalently,
\begin{align}
    \sigma_F
    &\simeq
    \frac{\sqrt{6a}}{\pi b T^{3/2}},
    \\
    \sigma_{\dot f_{\rm mid}}
    &\simeq
    \frac{\sqrt{360a}}{\pi b T^{5/2}},
    \\
    \sigma_{\ddot f}
    &\simeq
    \frac{\sqrt{50400a}}{\pi b T^{7/2}}.
\end{align}

This extension is useful because it separates the strongly correlated polynomial phase parameters into approximately independent local coordinates. A practical implementation need not perform a prohibitively dense three-dimensional search over $(f_{\rm start},\dot f,\ddot f)$. Instead, one may first localize the peak with a coarse grid in $\ddot f$ and then fit the local coherent response in the transformed coordinates $(F,\dot f_{\rm mid},\ddot f)$. Such a procedure would provide a computationally efficient route to estimating $\ddot f$ and its uncertainty, while retaining the same physical interpretation used for the two-parameter $(f,\dot f)$ case.

We have not used this extension in the present analysis. The formulae above should therefore be regarded as the natural analytic generalization of the local coherent-response method, rather than as a calibrated estimator for the present data set. A dedicated validation with Poisson simulations, different signal-to-noise ratios, observing gaps, pulse shapes, and source parameters is required before applying this three-parameter version to real event lists.

\subsection{Source-matched simulated event lists}
\label{app:source_matched_simulations}

The uncertainty calibration in the main text is based on simplified sinusoidal simulations with known injected rotational parameters. In addition to the idealized tests, we also generated source-matched simulations for the three AstroSat event lists analysed in this work. These simulations use the approximate count-rate scale, adopted rotational parameters, and good-time-interval structure of each observation. Their purpose is not to reproduce the detailed astrophysical variability of each source, but to verify that event lists with similar photon statistics and observing windows lead to recovered-parameter scatter consistent with the uncertainty estimates derived from the local $Z_1^2$ response.

\begin{figure*}
    \centering
    \includegraphics[width=0.9\linewidth]{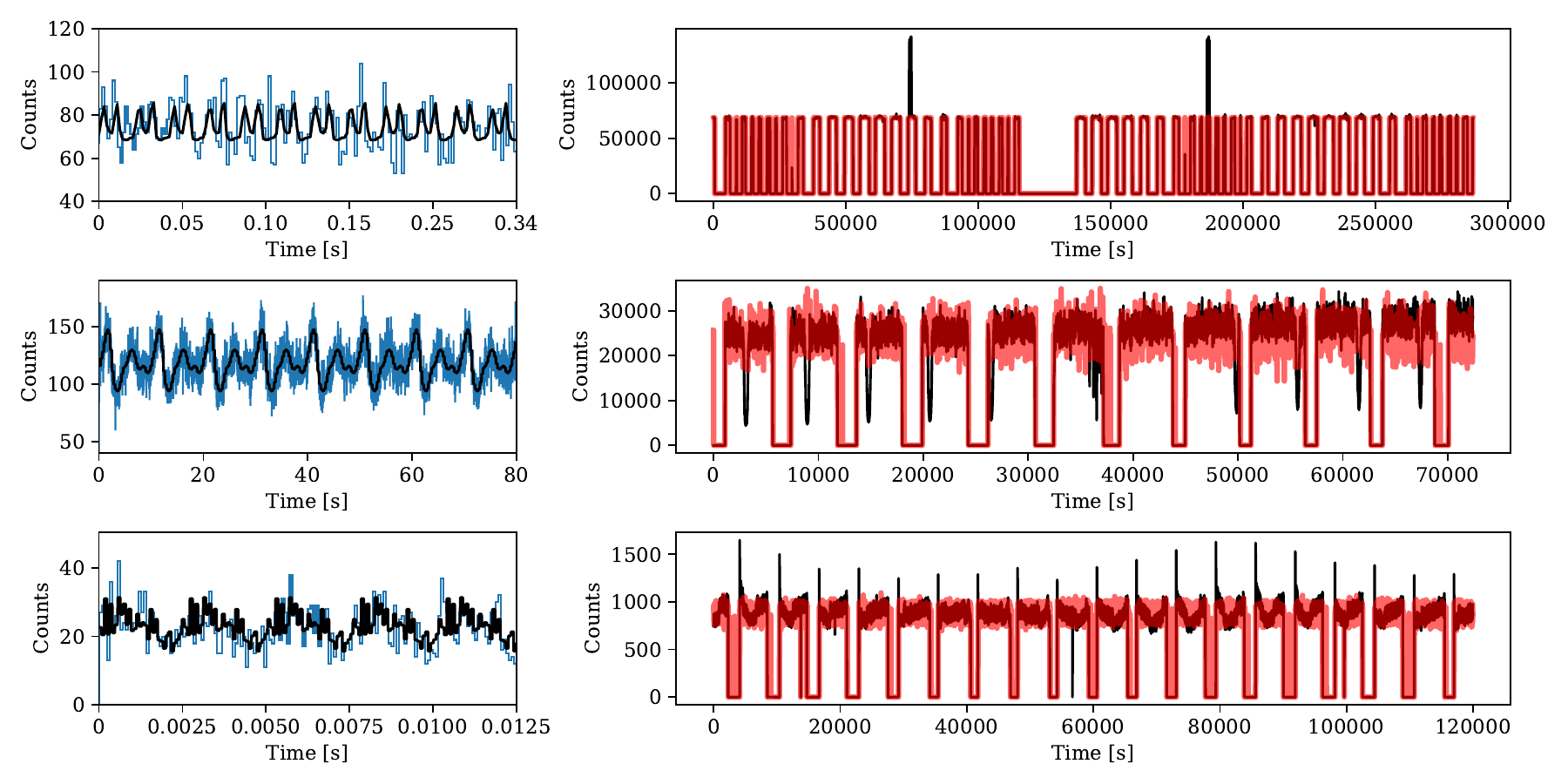}
    \caption{
    Source-matched simulated event lists used for additional validation of the timing-uncertainty estimates.
    The left column shows the input simulated light curves, with the corresponding binned simulated event data overlaid, for the Crab pulsar, SAX~J1808.4$-$3658, and Swift~J0243.6+6124 from top to bottom.
    The right column compares the binned AstroSat/LAXPC event lists with the corresponding simplified simulations.
    The simulations use source-specific choices of $a$, $b$, $f$, $\dot f$, and $T$, together with the good-time-interval structure of the observations, so that the overall count-rate scale and data gaps are approximately reproduced.
    These simulations are intended as controlled Poisson tests of the recovery procedure, not as detailed physical models of the source variability.
    }
    \label{fig:simulated_3}
\end{figure*}

Figure~\ref{fig:simulated_3} shows the observed and simulated time series for the Crab pulsar, SAX~J1808.4$-$3658, and Swift~J0243.6+6124. The figure is included here to document the observing-window structure, including data gaps, and to show the level at which the simplified simulations match the overall count-rate scale of the real event lists. The simulations should therefore be interpreted as controlled statistical tests of the timing-recovery procedure, rather than as physical models of the full source behaviour.

\bibliographystyle{mnras}
\bibliography{refs} 

@article{wijnands1998millisecond,
  title={A millisecond pulsar in an X-ray binary system},
  author={Wijnands, Rudy and Van Der Klis, Michiel},
  journal={Nature},
  volume={394},
  number={6691},
  pages={344--346},
  year={1998},
  publisher={Nature Publishing Group UK London}
}

@article{chakrabarty1998two,
  title={The two-hour orbit of a binary millisecond X-ray pulsar},
  author={Chakrabarty, Deepto and Morgan, Edward H},
  journal={Nature},
  volume={394},
  number={6691},
  pages={346--348},
  year={1998},
  publisher={Nature Publishing Group UK London}
}

@article{sharma2023astrosat,
  title={AstroSat observation of the accreting millisecond X-ray pulsar SAX J1808. 4--3658 during its 2019 outburst},
  author={Sharma, Rahul and Sanna, Andrea and Beri, Aru},
  journal={Monthly Notices of the Royal Astronomical Society},
  volume={519},
  number={3},
  pages={3811--3818},
  year={2023},
  publisher={Oxford University Press}
}

@article{beri2021astrosat,
  title={AstroSat observations of the first Galactic ULX pulsar Swift J0243. 6+ 6124},
  author={Beri, Aru and Naik, Sachindra and Singh, Kulinder Pal and Jaisawal, Gaurava K and Bhattacharyya, Sudip and Charles, Philip and Ho, Wynn CG and Maitra, Chandreyee and Bhattacharya, Dipankar and Dewangan, Gulab C and others},
  journal={Monthly Notices of the Royal Astronomical Society},
  volume={500},
  number={1},
  pages={565--575},
  year={2021},
  publisher={Oxford University Press}
}

@article{kennea2017swift,
  title={Swift J0243. 6+ 6124: Swift discovery of an accreting NS transient},
  author={Kennea, JA and Lien, AY and Krimm, HA and Cenko, SB and Siegel, MH},
  journal={The Astronomer's Telegram},
  volume={10809},
  pages={1},
  year={2017}
}

@article{cenko2017grb,
  title={GRB 171003A: Swift detection of a burst or a Galactic Transient.},
  author={Cenko, SB and Barthelmy, SD and D'Avanzo, P and Kennea, JA and Lien, AY and Marshall, FE and Palmer, DM and Siegel, MH and Tohuvavohu, A},
  journal={GRB Coordinates Network},
  volume={21960},
  pages={1},
  year={2017}
}

@article{jenke2017fermi,
  title={Fermi GBM detects pulsations from Swift J0243. 6+ 6124},
  author={Jenke, P and Wilson-Hodge, CA},
  journal={The Astronomer's Telegram},
  volume={10812},
  pages={1},
  year={2017}
}

@article{jaisawal2018understanding,
  title={Understanding the spectral and timing behaviour of a newly discovered transient X-ray pulsar Swift J0243. 6+ 6124},
  author={Jaisawal, Gaurava K and Naik, Sachindra and Chenevez, J{\'e}r{\^o}me},
  journal={Monthly Notices of the Royal Astronomical Society},
  volume={474},
  number={4},
  pages={4432--4437},
  year={2018},
  publisher={Oxford University Press}
}

@article{Lyne1993_CrabTiming,
  author  = {Lyne, A. G. and Pritchard, R. S. and Smith, F. G.},
  title   = {23 years of Crab pulsar rotational history},
  journal = {Monthly Notices of the Royal Astronomical Society},
  volume  = {265},
  pages   = {1003--1012},
  year    = {1993},
  doi     = {10.1093/mnras/265.4.1003}
}

@article{davies1990improved,
  title={An improved test for periodicity},
  author={Davies, SR},
  journal={Monthly Notices of the Royal Astronomical Society},
  volume={244},
  pages={93--95},
  year={1990}
}

@article{Buccheri1983,
  title={Search for pulsed gamma-ray emission from radio pulsars in the COS-B data},
  author={Buccheri, R and Bennett, K and Bignami, GF and Bloemen, JBGM and Boriakoff, V and Caraveo, PA and Hermsen, W and Kanbach, G and Manchester, RN and Masnou, JL and others},
  journal={Astronomy and Astrophysics},
  volume={128},
  pages={245--251},
  year={1983}
}

@ARTICLE{busching2010h,
  author  = {{B{\"u}sching}, I. and {de Jager}, O.~C.},
  title   = {{The H-test probability distribution revisited: Improved sensitivity}},
  journal = {\apj},
  year    = {2010},
  volume  = {725},
  number  = {1},
  pages   = {898--901},
  doi     = {10.1088/0004-637X/725/1/898}
}

@article{ransom2002fourier,
  title={Fourier techniques for very long astrophysical time-series analysis},
  author={Ransom, Scott M and Eikenberry, Stephen S and Middleditch, John},
  journal={The Astronomical Journal},
  volume={124},
  number={3},
  pages={1788},
  year={2002},
  publisher={IOP Publishing}
}

@article{Verbiest2009,
  author  = {Verbiest, J. P. W. and Bailes, M. and van Straten, W. and Hobbs, G. B. and others},
  title   = {Precision timing of PSR J0437–4715: an accurate pulsar distance, a high pulsar mass and a limit on the variation of Newton's gravitational constant},
  journal = {Monthly Notices of the Royal Astronomical Society},
  volume  = {400},
  number  = {2},
  pages   = {951--968},
  year    = {2009},
  doi     = {10.1111/j.1365-2966.2009.15508.x}
}

@article{huppenkothen2019stingray,
  title={Stingray: a modern Python library for spectral timing},
  author={Huppenkothen, Daniela and Bachetti, Matteo and Stevens, Abigail L and Migliari, Simone and Balm, Paul and Hammad, Omar and Khan, Usman Mahmood and Mishra, Himanshu and Rashid, Haroon and Sharma, Swapnil and others},
  journal={The Astrophysical Journal},
  volume={881},
  number={1},
  pages={39},
  year={2019},
  publisher={IOP Publishing}
}

@ARTICLE{2017ApJS..231...10A,
       author = {{Antia}, H.~M. and {Yadav}, J.~S. and {Agrawal}, P.~C. and others},
        title = "{Calibration of the Large Area X-Ray Proportional Counter (LAXPC) Instrument on board AstroSat}",
      journal = {\apjs},
         year = 2017,
        month = jul,
       volume = {231},
       number = {1},
          eid = {10},
        pages = {10},
          doi = {10.3847/1538-4365/aa7a0e},
archivePrefix = {arXiv},
       eprint = {1702.08624}
}

@ARTICLE{2020ApJ...898...38B,
       author = {{Bult}, Peter and {Chakrabarty}, Deepto and {Arzoumanian}, Zaven and {Gendreau}, Keith C. and {Guillot}, Sebastien and {Malacaria}, Christian and {Ray}, Paul. S. and {Strohmayer}, Tod E.},
        title = "{Timing the Pulsations of the Accreting Millisecond Pulsar SAX J1808.4-3658 during Its 2019 Outburst}",
      journal = {\apj},
         year = 2020,
        month = jul,
       volume = {898},
       number = {1},
          eid = {38},
        pages = {38},
          doi = {10.3847/1538-4357/ab9827},
archivePrefix = {arXiv},
       eprint = {1910.03062}
}

@INPROCEEDINGS{2014SPIE.9144E..1SS,
       author = {{Singh}, Kulinder Pal and {Tandon}, S.~N. and {Agrawal}, P.~C. and others},
        title = "{ASTROSAT mission}",
    booktitle = {Space Telescopes and Instrumentation 2014: Ultraviolet to Gamma Ray},
         year = 2014,
       editor = {{Takahashi}, Tadayuki and {den Herder}, Jan-Willem A. and {Bautz}, Mark},
       series = {Society of Photo-Optical Instrumentation Engineers (SPIE) Conference Series},
       volume = {9144},
        month = jul,
          eid = {91441S},
        pages = {91441S},
          doi = {10.1117/12.2062667}
}

@ARTICLE{Leahy1983b,
       author = {{Leahy}, D.~A. and {Elsner}, R.~F. and {Weisskopf}, M.~C.},
        title = "{On searches for periodic pulsed emission - The Rayleigh test compared to epoch folding}",
      journal = {\apj},
         year = 1983,
        month = sep,
       volume = {272},
        pages = {256-258},
          doi = {10.1086/161288}
}

@ARTICLE{2006AdSpR..38.2989A,
       author = {{Agrawal}, P.~C.},
        title = "{A broad spectral band Indian Astronomy satellite {\textquoteleft}Astrosat{\textquoteright}}",
      journal = {Advances in Space Research},
         year = 2006,
        month = jan,
       volume = {38},
       number = {12},
        pages = {2989-2994},
          doi = {10.1016/j.asr.2006.03.038}
}

@article{Hewish1968,
  author = {Hewish, A. and Bell, S. J. and Pilkington, J. D. H. and Scott, P. F. and Collins, R. A.},
  title = {Observation of a Rapidly Pulsating Radio Source},
  journal = {Nature},
  year = {1968},
  volume = {217},
  pages = {709--713},
  doi = {10.1038/217709a0}
}

@article{Lattimer2004,
  author = {Lattimer, J. M. and Prakash, M.},
  title = {The physics of neutron stars},
  journal = {Science},
  year = {2004},
  volume = {304},
  pages = {536--542},
  doi = {10.1126/science.1090720}
}

@article{Kramer2006,
  author = {Kramer, M. and Stairs, I. H. and Manchester, R. N. and others},
  title = {Tests of General Relativity from Timing the Double Pulsar},
  journal = {Science},
  year = {2006},
  volume = {314},
  pages = {97--102},
  doi = {10.1126/science.1132305}
}

@book{Manchester1977,
  author = {Manchester, R. N. and Taylor, J. H.},
  title = {Pulsars},
  publisher = {W. H. Freeman},
  year = {1977},
  address = {San Francisco}
}

@book{Lorimer2005,
  author = {Lorimer, D. R. and Kramer, M.},
  title = {Handbook of Pulsar Astronomy},
  publisher = {Cambridge University Press},
  year = {2005},
  address = {Cambridge}
}

@article{Agazie2023,
  author = {Agazie, G. and others},
  title = {The NANOGrav 15 yr Data Set: Evidence for a Gravitational-wave Background},
  journal = {The Astrophysical Journal Letters},
  year = {2023},
  volume = {951},
  pages = {L8},
  doi = {10.3847/2041-8213/acdac6}
}

@article{Antoniadis2023,
  author = {Antoniadis, J. and others},
  title = {The second data release from the European Pulsar Timing Array III. Search for gravitational wave signals},
  journal = {Astronomy \& Astrophysics},
  year = {2023},
  volume = {678},
  pages = {A50},
  doi = {10.1051/0004-6361/202346844}
}

@article{Taylor1992,
  author = {Taylor, J. H.},
  title = {Pulsar Timing and Relativistic Gravity},
  journal = {Philosophical Transactions of the Royal Society A},
  year = {1992},
  volume = {341},
  pages = {117--134},
  doi = {10.1098/rsta.1992.0088}
}

@article{Hobbs2006,
  author = {Hobbs, G. B. and Edwards, R. T. and Manchester, R. N.},
  title = {TEMPO2, a new pulsar-timing package - I. Overview},
  journal = {Monthly Notices of the Royal Astronomical Society},
  year = {2006},
  volume = {369},
  pages = {655--672},
  doi = {10.1111/j.1365-2966.2006.10302.x}
}

@ARTICLE{Ray2011,
       author = {{Ray}, P.~S. and {Kerr}, M. and {Parent}, D. and {Abdo}, A.~A. and {Guillemot}, L. and {Ransom}, S.~M. and {Rea}, N. and {Wolff}, M.~T. and {Makeev}, A. and {Roberts}, M.~S.~E. and {Camilo}, F. and {Dormody}, M. and {Freire}, P.~C.~C. and {Grove}, J.~E. and {Gwon}, C. and {Harding}, A.~K. and {Johnston}, S. and {Keith}, M. and {Kramer}, M. and {Michelson}, P.~F. and {Romani}, R.~W. and {Saz Parkinson}, P.~M. and {Thompson}, D.~J. and {Weltevrede}, P. and {Wood}, K.~S. and {Ziegler}, M.},
        title = "{Precise {\ensuremath{\gamma}}-ray Timing and Radio Observations of 17 Fermi {\ensuremath{\gamma}}-ray Pulsars}",
      journal = {\apjs},
         year = 2011,
        month = jun,
       volume = {194},
       number = {2},
          eid = {17},
        pages = {17},
          doi = {10.1088/0067-0049/194/2/17},
archivePrefix = {arXiv},
       eprint = {1011.2468},
 primaryClass = {astro-ph.HE},
       adsurl = {https://ui.adsabs.harvard.edu/abs/2011ApJS..194...17R}
}

@article{Kerr2011,
  author = {Kerr, M.},
  title = {Maximum Likelihood Analysis of Pulsar Light Curves},
  journal = {The Astrophysical Journal},
  year = {2011},
  volume = {732},
  pages = {38},
  doi = {10.1088/0004-637X/732/1/38}
}

@article{Atwood2009,
  author = {{Atwood}, W.~B. and {Abdo}, A.~A. and {Ackermann}, M. and others},
  title = "{The Large Area Telescope on the Fermi Gamma-ray Space Telescope Mission}",
  journal = {\apj},
  year = 2009,
  volume = 697,
  pages = {1071-1102},
  doi = {10.1088/0004-637X/697/2/1071}
}

@inproceedings{Gendreau2016,
  author = {{Gendreau}, K.~C. and {Arzoumanian}, Z. and {Adkins}, P.~W. and others},
  title = "{The Neutron star Interior Composition Explorer (NICER): design and development}",
  booktitle = {Space Telescopes and Instrumentation 2016: Ultraviolet to Gamma Ray},
  series = {SPIE Conference Series},
  volume = 9905,
  year = 2016,
  eid = {99051H},
  doi = {10.1117/12.2231304}
}

@ARTICLE{1978ApJ...224..953S,
       author = {{Stellingwerf}, R.~F.},
        title = "{Period determination using phase dispersion minimization.}",
      journal = {\apj},
         year = 1978,
        month = sep,
       volume = {224},
        pages = {953-960},
          doi = {10.1086/156444}
}

@article{Rayleigh1919,
  author    = {{Lord Rayleigh}},
  title     = {XXXI. On the problem of random vibrations, and of random flights in one, two, or three dimensions},
  journal   = {The London, Edinburgh, and Dublin Philosophical Magazine and Journal of Science (Series 6)},
  year      = {1919},
  volume    = {37},
  number    = {220},
  pages     = {321--347},
  publisher = {Taylor \& Francis},
  doi       = {10.1080/14786440408635894}
}

@ARTICLE{2018AandA...613A..19D,
       author = {{Doroshenko}, V. and {Tsygankov}, S. and {Santangelo}, A.},
        title = "{Orbit and intrinsic spin-up of the newly discovered transient X-ray pulsar Swift J0243.6+6124}",
      journal = {\aap},
         year = 2018,
        month = may,
       volume = {613},
          eid = {A19},
        pages = {A19},
          doi = {10.1051/0004-6361/201732208},
archivePrefix = {arXiv},
       eprint = {1710.10912}
}

@ARTICLE{Leahy1987,
       author = {{Leahy}, D.~A.},
        title = "{Searches for pulsed emission with application to four globular cluster X-ray sources}",
      journal = {\aap},
         year = 1987,
        month = jun,
       volume = {180},
        pages = {275-277}
}

@ARTICLE{2005ApJ...618..866N,
       author = {{Naik}, Sachindra and {Paul}, Biswajit and {Callanan}, Paul J.},
        title = "{X-ray timing of the accretion-powered millisecond pulsar XTE J1814-338}",
      journal = {\apj},
         year = 2005,
        month = jan,
       volume = {618},
       number = {2},
        pages = {866-870},
          doi = {10.1086/426065}
}

@ARTICLE{2017NewA...56...94B,
       author = {{Bhattacharyya}, S. and others}, 
        title = "{A timing and spectral study of the transient X-ray pulsar...}", 
      journal = {\na},
         year = 2017,
        month = oct,
       volume = {56},
        pages = {94-101}
}

@BOOK{Bloomfield1976,
       author = {{Bloomfield}, Peter},
        title = "{Fourier Analysis of Time Series: An Introduction}",
    publisher = {John Wiley \& Sons, New York},
         year = 1976}

@ARTICLE{Kovacs1980,
       author = {{Kovacs}, G.},
        title = "{Fast period-searching in unevenly sampled observational data}",
      journal = {\apss},
         year = 1981,
        month = aug,
       volume = {78},
       number = {1},
        pages = {175-190},
          doi = {10.1007/BF00648943}
}

@ARTICLE{Larsson1996,
       author = {{Larsson}, S.},
        title = "{Parameter estimation in periodograms.}",
      journal = {\aaps},
         year = 1996,
        month = jun,
       volume = {117},
        pages = {197-201}
}

@ARTICLE{Edwards2006,
       author = {{Edwards}, R.~T. and {Hobbs}, G.~B. and {Manchester}, R.~N.},
        title = "{TEMPO2, a new pulsar-timing package - II. The timing model and precision estimates}",
      journal = {\mnras},
         year = 2006,
        month = nov,
       volume = {372},
       number = {4},
        pages = {1549-1574},
          doi = {10.1111/j.1365-2966.2006.10870.x}
}

@BOOK{Bretthorst1988,
       author = {{Bretthorst}, G. Larry},
        title = "{Bayesian Spectrum Analysis and Parameter Estimation}",
    publisher = {Springer-Verlag, Berlin},
       series = {Lecture Notes in Statistics},
       volume = {48},
         year = 1988,
          doi = {10.1007/978-1-4757-4343-2}
}

@ARTICLE{chang,

       author = {{Chang}, Chulhoon and {Pavlov}, George G. and {Kargaltsev}, Oleg and {Shibanov}, Yurii A.},

        title = "{X-Ray Observations of the Young Pulsar J1357{\textemdash}6429 and Its Pulsar Wind Nebula}",

      journal = {\apj},

         year = 2012,

        month = jan,

       volume = {744},

       number = {2},

          eid = {81},

        pages = {81},

          doi = {10.1088/0004-637X/744/2/81},

archivePrefix = {arXiv},

       eprint = {1107.1819},

 primaryClass = {astro-ph.HE},

       adsurl = {https://ui.adsabs.harvard.edu/abs/2012ApJ...744...81C}

}

@ARTICLE{Boldin2013,
       author = {{Boldin}, P.~A. and {Tsygankov}, S.~S. and {Lutovinov}, A.~A.},
        title = "{Timing of X-ray pulsars: Evolution of the pulsation period and the cyclotron line energy}",
      journal = {Astronomy Letters},
         year = 2013,
        month = jun,
       volume = {39},
       number = {6},
        pages = {375-385},
          doi = {10.1134/S1063773713060029}
}

@ARTICLE{2023arXiv231106620S,
       author = {{Singhal}, Akshat and {Jain}, Ishan and {Bala}, Suman and {Bhalerao}, Varun},
        title = "{Quantifying Period Uncertainty in X-ray Pulsars with Poisson-Limited Data}",
      journal = {arXiv e-prints},
         year = 2023,
        month = nov,
          eid = {arXiv:2311.06620},
        pages = {arXiv:2311.06620},
          doi = {10.48550/arXiv.2311.06620},
archivePrefix = {arXiv},
       eprint = {2311.06620},
 primaryClass = {astro-ph.HE},
       adsurl = {https://ui.adsabs.harvard.edu/abs/2023arXiv231106620S}
}

@article{Shaltev2014,
  author       = {Shaltev, Miroslav and Leaci, Paola and Papa, Maria Alessandra and Prix, Reinhard},
  title        = {Fully coherent follow-up of continuous gravitational-wave candidates},
  journal      = {Physical Review D},
  volume       = {89},
  pages        = {124030},
  year         = {2014},
  doi          = {10.1103/PhysRevD.89.124030},
  eprint       = {1303.2471},
  archivePrefix= {arXiv},
  primaryClass = {gr-qc}
}

@article{Piccinni2019,
  author       = {Piccinni, O. J. and Frasca, S. and Astone, P. and D'Antonio, S. and Intini, G. and Leaci, P. and Mastrogiovanni, S. and Miller, A. and Palomba, C. and Singhal, A.},
  title        = {A new data analysis framework for the search of continuous gravitational wave signals},
  journal      = {Classical and Quantum Gravity},
  volume       = {36},
  number       = {1},
  pages        = {015008},
  year         = {2019},
  doi          = {10.1088/1361-6382/aaefb5},
  eprint       = {1811.04730},
  archivePrefix= {arXiv},
  primaryClass = {gr-qc}
}

@ARTICLE{Singhal2019,
       author = {{Singhal}, A. and {Leaci}, P. and {Astone}, P. and {D'Antonio}, S. and {Frasca}, S. and {Intini}, G. and {La Rosa}, I. and {Mastrogiovanni}, S. and {Miller}, A. and {Muciaccia}, F. and {Palomba}, C. and {Piccinni}, O.},
        title = "{A resampling algorithm to detect continuous gravitational-wave signals from neutron stars in binary systems}",
      journal = {Classical and Quantum Gravity},
         year = 2019,
        month = oct,
       volume = {36},
       number = {20},
          eid = {205015},
        pages = {205015},
          doi = {10.1088/1361-6382/ab4367},
       adsurl = {https://ui.adsabs.harvard.edu/abs/2019CQGra..36t5015S}
}

@ARTICLE{Raman2016,  author        = {{Raman}, Gayathri and {Paul}, Biswajit and {Bhattacharya}, Dipankar and {Mohan}, Vijay},  title         = {{SALT observation of X-ray pulse reprocessing in 4U 1626-67}},  journal       = {\mnras},  year          = {2016},  volume        = {458},  number        = {2},  pages         = {1302--1310},  doi           = {10.1093/mnras/stw290},  archivePrefix = {arXiv},  eprint        = {1602.01607},  primaryClass  = {astro-ph.HE},  adsurl        = {https://ui.adsabs.harvard.edu/abs/2016MNRAS.458.1302R}}

@ARTICLE{Raman2021,  author  = {{Raman}, Gayathri and {Varun} and {Paul}, Biswajit and {Bhattacharya}, Dipankar},  title   = {{AstroSat detection of a mHz quasi-periodic oscillation and cyclotron line in IGR J19294+1816 during the 2019 outburst}},  journal = {\mnras},  year    = {2021},  volume  = {508},  number  = {4},  pages   = {5578--5586},  doi     = {10.1093/mnras/stab2835},  adsurl  = {https://ui.adsabs.harvard.edu/abs/2021MNRAS.508.5578R}}





\label{lastpage}
\end{document}